\documentclass[preprint2]{proto}
\usepackage{times}
\newcommand{\refs}{\par\noindent\hangindent=1pc\hangafter=1}
\def\I{IRAS\,20126+4104}
\def\AMM{NH$_3$}
\def\COI{\mbox{$^{13}$CO}}
\def\MCN{\mbox{CH$_3$CN}}
\def\WAT{H$_2$O}
\def\METH{CH$_3$OH}
\def\CSI{C$^{34}$S}
\def\HII{H{\sc ii}}
\def\kms{\mbox{km~s$^{-1}$}}

\def\Mst{\mbox{$M_\star$}}
\def\Msun{\mbox{$M_\odot$}}
\def\Lsun{\mbox{$L_\odot$}}
\def\My{\mbox{$M_\odot$yr$^{-1}$}}

\def\la{\mathrel{\mathchoice {\vcenter{\offinterlineskip\halign{\hfil
$\displaystyle##$\hfil\cr<\cr\sim\cr}}}
{\vcenter{\offinterlineskip\halign{\hfil$\textstyle##$\hfil\cr
<\cr\sim\cr}}}
{\vcenter{\offinterlineskip\halign{\hfil$\scriptstyle##$\hfil\cr
<\cr\sim\cr}}}
{\vcenter{\offinterlineskip\halign{\hfil$\scriptscriptstyle##$\hfil\cr
<\cr\sim\cr}}}}}
\def\ga{\mathrel{\mathchoice {\vcenter{\offinterlineskip\halign{\hfil
$\displaystyle##$\hfil\cr>\cr\sim\cr}}}
{\vcenter{\offinterlineskip\halign{\hfil$\textstyle##$\hfil\cr
>\cr\sim\cr}}}
{\vcenter{\offinterlineskip\halign{\hfil$\scriptstyle##$\hfil\cr
>\cr\sim\cr}}}
{\vcenter{\offinterlineskip\halign{\hfil$\scriptscriptstyle##$\hfil\cr
>\cr\sim\cr}}}}}

\begin{document}

\title{\textbf{\LARGE Disks around Young O-B (Proto)Stars: Observations and Theory}}

\author {\textbf{\large R. Cesaroni, D. Galli}}
\affil{\small\em INAF -- Osservatorio Astrofisico di Arcetri}
\author {\textbf{\large G. Lodato}}
\affil{\small\em Institute of Astronomy}
\author {\textbf{\large C.M. Walmsley}}
\affil{\small\em INAF -- Osservatorio Astrofisico di Arcetri}
\author {\textbf{\large Q. Zhang}}
\affil{\small\em Harvard-Smithsonian Center for Astrophysics}


\begin{abstract}
\baselineskip = 11pt
\leftskip = 0.65in 
\rightskip = 0.65in
\parindent=1pc
{\small 
Disks are a natural outcome of the star formation process in which
they play a crucial role. Luminous, massive stars of spectral type
earlier than B4 are likely to be those that benefit most from the
existence of accretion disks, which may significantly reduce the effect of
radiation pressure on the accreting material. The scope of the present
contribution is to review the current knowledge about disks in young high-mass
(proto)stars and discuss their implications. The
issues of disk stability and lifetime are also discussed.
We conclude that for protostars of less than $\sim$20~\Msun, disks with mass
comparable to that of the central star are common. Above this limit the
situation is unclear and there are no good examples of proto O4--O8 stars
surrounded by accretion disks: in these objects only huge, massive, toroidal,
non-equilibrium rotating structures are seen. It is clear on the other hand
that the observed disks in stars of 10--20~\Msun\ are likely to be unstable
and with short lifetimes.
 \\~\\~\\~}

\end{abstract}

\section{\textbf{INTRODUCTION}}

Disk formation seems to be the natural consequence of star formation.
For low-mass stars, this is attested to by the existence of
pre-main sequence (PMS) star disks
as shown by a number of publications including several reviews in
Protostars and Planets IV
(see, e.g.,
{\it Mathieu et al.}, 2000; {\it O'Dell and Wen}, 1994; {\it O'Dell}, 2001; {\it Guilloteau
et al.}, 1999; {\it Simon et al.}, 2000; {\it Mundy et al.}, 2000; {\it Calvet et al.},
2000; {\it Hollenbach et al.}, 2000; {\it Wilner and Lay}, 2000).
Noticeably, all of these refer only to disks in low- to intermediate-mass
protostars, while the topic of disks in O-B stars is reviewed
for the first time in the present contribution.
Millimeter observations of disks around young early B stars ({\it Natta et
al.}, 2000; {\it Fuente et al.}, 2003) give upper limits (or in one case a
detection) less than 0.001 times the mass of the star, in contrast to PMS
stars of B5 and later.  Thus the evidence is that O and
early B stars lose any disks they may originally have had prior to
becoming optically visible after roughly $10^6$~yr. This article aims
at discussing the question of what happened in those first million
years.
That in turn depends on understanding what happens in the embedded
phase where we depend on  millimeter and infrared measurements to
detect disks. Before getting to this, it is worth noting that
there is evidence for disks around what one presumes to be embedded
young O-B stars with characteristics similar to those of the BN object
in Orion (e.g., {\it Scoville et al.}, 1983; {\it Bik and Thi}, 2004).  This comes
from high resolution spectra of the CO overtone bands and
suggests that the lines are emitted in a disk a few AU from the star at
temperatures above 1500~K.  Note that disks  of this
size would not be detectable by the mm observations mentioned earlier.
Nevertheless, it seems likely that also disks on AU scales disappear rapidly.
{\it Hollenbach et al.} (2000)
summarize disk dispersal mechanisms and many of these are more
effective in the high pressure cores where O-B stars form.  Stellar
wind stripping and photo-evaporation occurs naturally in the
neighbourhood of hot massive stars whereas disks on AU scales
dissipate rapidly due to viscous accretion.

On the other hand, massive
stars have to form and one may need to accrete through a disk in order
to form them.
This is mainly because the most obvious barrier to massive star
formation is the action of radiation pressure on infalling gas and
dust (e.g., {\it Kahn}, 1974; {\it Wolfire and Cassinelli}, 1987). Current models
({\it Jijina and Adams}, 1996; {\it Stahler et al.}, 2000; {\it Yorke and Sonnhalter}, 2002;
{\it Yorke}, 2004a; {\it Krumholz et al.}, 2005) suggest that the most effective way
of overcoming radiation pressure is to accrete via a disk.
In this way, however, forming a star requires loss of angular momentum and one way
of doing this is via an outflow which may be powered by a wind
({\it Shu et al.}, 2000; {\it K\"onigl and Pudritz}, 2000). As discussed in
Section~2.1, there is evidence for outflows from young massive
(proto)stars ({\it Richer et al.}, 2000) and thus implicitly evidence for
disks. Thus it seems likely that disks are an essential part of
high-mass star formation but that they are short lived and hence, by
inference, high-mass star formation is a rapid process.  In fact,
current models ({\it McKee and Tan}, 2003; see also the chapter by {\it Beuther et al.})
of the formation of high-mass
stars in clusters suggest  timescales  of the order of $10^5$ years and
accretion rates upwards of $10^{-4}$~\My.

One difficulty in understanding the observations of massive protostars
and their associated disks is our ignorance of the
evolution of the central star itself.  One can imagine scenarios where
the accreting protostar evolves rapidly to the zero-age main sequence (ZAMS)
and others where
the protostellar radius remains larger than the ZAMS value (and thus
the effective temperature smaller) for a protracted period of time
(e.g., {\it Nakano et al.}, 2000). In this latter case, the accretion luminosity
is a considerable fraction of the bolometric luminosity. This
alleviates somewhat the radiation pressure problem and postpones the
photo-evaporation of the disk. It also may explain why some luminous
young stellar objects (YSOs) appear to have a very low Lyman-continuum output
as measured by the observed radio free-free emission (e.g., {\it Molinari et al.},
1998; {\it Sridharan et al.}, 2002). In any case, the uncertainty about the
central star properties gives rise to an additional uncertainty in our
discussion of the surrounding disk.

Since it is useful to be guided in the first place by the
observations, we in Section~2 of this review summarize the observations
of disks around massive protostars, where by massive we mean
luminosities larger than $\sim10^3$~\Lsun, corresponding on the ZAMS to a
B4 spectral type ($\sim$8~\Msun).
This is roughly the point where the accretion
(free-fall) timescale equals the Kelvin-Helmholtz timescale (e.g., {\it Palla
and Stahler}, 1993; {\it Stahler et al.}, 2000). Conversely, all stars below
$\sim$8~\Msun\
will be called hereafter ``low-mass'', so that
we will {\it not} use the expression ``intermediate-mass'',
commonly adopted in the literature for stars in the range 2--8~\Msun.
It is worth noting that the threshold of 8~\Msun\ is not to be taken at face
value: this corresponds to an accretion rate of $10^{-5}$~\Msun~yr$^{-1}$,
while for higher or lower rates the critical mass may change significantly
({\it Palla and Stahler}, 1993). We will thus consider as massive also
(proto)stars with slightly lower masses than 8~\Msun.

In Section~3, we summarize the data for the well studied ``disk'' around the
protostar IRAS\,20126+4104, which we take to be our prototype.
In Sections~4 to~6, we discuss the evolution,
stability, and lifetime of massive disks around massive stars. We illustrate
the difficulties with classical disk theory to explain the accretion
rates needed to form massive stars on short timescales and that thus
``unconventional approaches'' (perhaps involving companions) might be
required. In Section~7, we give a brief summary and draw our conclusions.

\bigskip

\centerline{\textbf{ 2. OBSERVATIONAL ASPECTS}}
\bigskip

In this section we illustrate the techniques adopted to search for
disks in regions of high-mass star formation and the results
obtained, with special attention to the physical properties of the disks.
Eventually, we will draw some conclusions about the apparent scarcity
of detections, in an attempt to decide whether this is due to instrumental
limitations or instead related to the formation mechanism.

As already mentioned, the first step to undertake is the definition of
``disk''. Theoretically a disk may be defined as a long-lived, flat,
rotating structure in centrifugal equilibrium. Although not
easy to fulfill on an observational ground, this criterion may inspire a more
``pragmatical'' definition.

A morphological definition based on the disk geometry may not only be
difficult but even ambiguous, given that massive YSOs (and hence their disks)
are deeply embedded in the parental clumps. This makes the distinction
between disk and surrounding envelope hard to establish, even at
sub-arcsecond resolution. Disks could be identified from their spectral
energy distributions (SEDs), but such an identification would be model
dependent. The most reliable criterion is based on
the velocity field of the rotating disk: an edge-on
disk should be seen as a linear structure with a systematic velocity shift
along it. However, velocity gradients may be determined also by
infalling or outflowing gas, not only by rotation. To discriminate among the
these, one possibility is to take profit of the fact that
disks are probably always associated with large scale jets/outflows, ejected
along their rotation axes: this is likely true also for disks in high-mass
YSOs, as the disk-outflow association is well established in a large variety
of environments ranging from disks in low-mass protostars (e.g., {\it Burrows et
al.}, 1996) to those in active galactic nuclei (e.g., {\it Krolik}, 1999).  One may
hence compare the direction of the velocity gradient detected on a small
scale ($\la$0.1~pc), which traces the putative disk, to that of the
molecular flow (seen over $\la$1~pc): if the two are parallel, the disk
hypothesis can be ruled out, whereas perpendicularity supports the disk
interpretation.

In conclusion, one may use as indicative of the presence of a disk the
existence of a small ($\la$0.1~pc) molecular core, located at the geometrical
center of a bipolar outflow, with a velocity gradient perpendicular to the
outflow axis. This ``observer's definition'' is to be taken as a necessary
condition: a satisfactory disk identification not only needs more robust
observational evidence, but must satisfy theoretical constraints
related to disk stability and life-time. These will be discussed in Sections~4
to~6.

The association between disks and outflows is very important.  The fact that
outflows turn out to be common in star forming regions of all masses
({\it Shepherd}, 2003) suggests indirectly that disks exist also in massive YSOs.
The outflow detection rate also allows one to obtain a quantitative estimate
of the disk number in such regions. This is discussed in the following
section.

\bigskip
\noindent
\textbf{ 2.1 The outflow-disk connection and the expected number of disks}
\bigskip

During the last decade a number of systematic
searches for molecular outflows have been
carried out towards a variety of candidate high-mass YSOs. After the work by
{\it Shepherd and Churchwell} (1996a, 1996b) who searched for CO(1--0) line wings
in a sample of ultracompact (UC) \HII\ regions, subsequent surveys were made
by {\it Osterloh et al.} (1997), {\it Beuther et al.} (2002a) and {\it Zhang et al.} (2001,
2005), towards IRAS selected sources, and {\it Codella et al.} (2004) in
UC~\HII\ regions and maser sources. The estimated
outflow detection rate ranges from
39\% to 90\%, demonstrating that the phenomenon is ubiquitous in massive star
forming regions (SFRs). The parameters of these flows, such as mass,
momentum, and mass loss rate are at least an order of magnitude greater than
those of outflows from low-mass YSOs.  Nevertheless, this fact by itself does
not demonstrate that the massive outflows are powered by massive YSOs, but
they might arise from a cluster of low-mass stars. Based on knowledge of the
properties of outflows from low-mass YSOs and assuming a Salpeter initial
mass function, {\it Zhang et al.} (2005) have actually shown that a collection of
outflows from a cluster of low-mass stars may explain the total outflow mass
observed. However, the same authors conclude that this scenario is unlikely,
as a randomly oriented sample of outflows should not produce a clear
bipolarity such as that observed in massive flows.

In conclusion, single-dish observations indicate that outflows from
massive YSOs are common. This result is confirmed by high-angular resolution
observations of a limited number of objects (see, e.g., {\it Mart\'{\i} et al.}, 1993;
{\it Cesaroni et al.}, 1997, 1999a; {\it Hunter et al.}, 1999; {\it Shepherd et al.}, 2000; {\it Beuther
et al.}, 2002b, 2003, 2004a; {\it Fontani et al.}, 2004; {\it Su et al.}, 2004) where
interferometric observations have resolved the structure of several outflows,
proving that their properties are indeed different from those of flows in
low-mass SFRs.

An important consequence is that disks must also be widespread in massive
SFRs, if the (somewhat arbitrary) assumption that disks and outflows are
strictly associated is correct. Only direct observations of the circumstellar
environment on a sub-arcsecond scale may prove this conclusion.
In the next section we will illustrate the methods used to search for
circumstellar disks in massive YSOs.

\bigskip
\noindent
\textbf{ 2.2 The search for disks in massive YSOs}
\bigskip

Identifying candidate disks in massive (proto)stars requires careful
selection of targets and tracers, to overcome the problems related
to observations of massive YSOs. The most important of these are the large
distance (a few kpc, i.e. $\sim$10 times those to low-mass objects) and
richness of the environment (massive stars form in clusters), which often
complicates the interpretation of the results. Consequently, sensitive and
high angular resolution observations are needed. Albeit difficult to
establish during the protostellar phase, the luminosities of massive YSOs
must be significantly larger than those of low-mass stars, so that a
reasonable lower limit to search for massive YSOs is $\sim10^3$~\Lsun.
In an attempt to bias the search towards young sources, additional criteria
may be applied, such as association with maser sources (mainly water and
methanol), non-detection of free-free emission from associated \HII\ regions,
and/or constraints on the IRAS colours to filter out more evolved
objects.  Finally, the presence of molecular outflows may be used to identify
sources which are potentially more suited to host disks, for the reason
discussed in the previous section. Examples of catalogues of massive
protostellar candidates selected according to the previous criteria are those
by {\it Churchwell et al.} (1990), {\it Plume et al.} (1992), {\it Tofani et al.} (1995),
{\it Molinari et al.} (1996), {\it Sridharan et al.} (2002).

As for the choice of tracers, disks
are located in the densest and hottest
part of the molecular clump. It is hence necessary to look for optically
thin, high-temperature tracers, such as continuum emission at (sub)mm
wavelengths and line emission from high-energy levels of low-abundance
molecular species. In Table~1, we summarize the main techniques
used to search for disks in massive YSOs and a number of relevant references.
These may be classified in four categories depending on the tracer adopted.

\begin{deluxetable}{l|l}
\tabletypesize{\small}
\tablecaption{List of tracers used to search for disks in high-mass YSOs}
\tablewidth{0pt}
\tablehead{Tracer & References \\ }
\startdata
CH$_3$OH & {\it Norris et al.} (1998); {\it Phillips et al.} (1998);  \\
masers   & {\it Minier et al.} (1998, 2000); {\it Pestalozzi et al.} (2004) \\
         & {\it Edris et al.} (2005) \\
OH masers & {\it Hutawarakorn and Cohen} (1999); {\it Edris et al.} (2005) \\
SiO masers & {\it Barvainis} (1984); {\it Wright et al.} (1995) \\
           & {\it Greenhill et al.} (2004) \\
H$_2$O masers & {\it Torrelles et al.} (1998) ; \\
              & {\it Shepherd and Kurtz} (1999) \\
\hline
IR, mm, cm & {\it Yao et al.} (2000); {\it Shepherd et al.} (2001); \\
continuum  & {\it Preibisch et al.} (2003); {\it Gibb et al.} (2004a); \\
           & {\it Chini et al.} (2004); {\it Sridharan et al.} (2005); \\
	   & {\it Jiang et al.} (2005); {\it Puga et al.} (submitted) \\
\hline
NH$_3$, C$^{18}$O, & {\it Keto et al.} (1988); \\
CS, C$^{34}$S, 	   & {\it Cesaroni et al.} (1994, 1997, 1998, 1999a, 2005); \\
CH$_3$CN,          & {\it Zhang et al.} (1998a, 1998b, 2002); \\
HCOOCH$_3$	   & {\it Shepherd and Kurtz} (1999); {\it Olmi et al.} (2003); \\
		   & {\it Sandell et al.} (2003); {\it Gibb et al.} (2004b); \\
                   & {\it Beltr\'an et al.} (2004, 2005); \\
                   & {\it Beuther et al.} (2004b, 2005) \\
\enddata
\end{deluxetable}

\bigskip
\noindent
{\em 2.2.1 Continuum emission.}
 The dust in the disk is bound to emit as a grey body, characterized by
 temperatures from a few 10~K to several $\sim$100~K.
 This means that the continuum
 emission must peak at far-IR wavelengths. However,
 at present, (sub)arcsecond imaging is possible only in the (sub)millimeter and mid-IR regimes, due to
 instrumental
 limitations and poor atmospheric transparency which make (sub)arcsecond
 imaging impossible in the far-IR.
 Multi-wavelength information is nevertheless helpful to
 identify the possible contribution from ionized stellar winds to the
 continuum emission, which may be confused with dust (i.e. disk) emission
 especially at 7~mm (see, e.g., {\it Gibb et al.}, 2004a). The major problem with
 continuum imaging is related to the fact that massive young (proto)stars,
 unlike their low-mass counterparts, are still deeply embedded in their
 parental clumps: this makes it very difficult to decouple the disk from the
 surrounding envelope, even at sub-arcsecond resolutions. The problem may be
 less critical in the near-IR, where the large extinction in the plane of the
 disk makes it possible to see the disk (if close to edge on) as a dark
 silhouette against the bright background. Examples of this are found in M17
 ({\it Chini et al.}, 2004), \I\ ({\it Sridharan et al.}, 2005), and
 G5.89--0.39 ({\it Puga et al.}, submitted).  Also, near-IR polarimetric
 imaging may be helpful in some cases ({\it Yao et al.}, 2000; {\it Jiang et
 al.}, 2005).  Notwithstanding a few encouraging results, continuum
 observations alone can hardly achieve convincing evidence of the presence of
 disks: line observations are needed in addition and are necessary to study
 the rotation velocity field.

\bigskip
\noindent
{\em 2.2.2 Maser lines.} Maser emission is concentrated in narrow ($\la$1~\kms),
 strong (up to $10^6$~Jy) lines and arises from regions (``spots'') which may
 be as small as a few AU (e.g., {\it Elitzur}, 1992). These characteristics make it
 suitable for milli-arcsecond resolution studies, which can investigate the
 distribution, line of sight velocities, and even proper motions of the maser
 ``spots''. One may thus obtain a detailed picture of the kinematics of the
 circumstellar environment on scales (a few 10~AU) not accessible by any
 other means. The shortcoming of this technique is that the strength of maser
 lines is very sensitive to a variety of factors (amplification path,
 inversion conditions, velocity gradients, etc.), which make it very hard if
 not impossible to derive useful physical parameters (temperature, density)
 from the observables.  As shown in Table~1, a large number of
 tentative disk detections are reported in the literature using maser
 observations. In all cases the authors find a linear distribution of maser
 spots with velocity changing systematically along it. This is taken as an
 indication of rotation in an edge-on disk.  As discussed at the beginning of
 Section~2, such an evidence must be corroborated by the presence of a
 jet/outflow perpendicular to the maser distribution, which is not always the
 case. It hence remains unclear whether the velocity gradients observed are
 related to rotation, infall, or expanding motions. In fact, the
 interpretation of CH$_3$OH masers as disks rotating about massive YSOs (see,
 e.g., {\it Norris et al.}, 1998; {\it Pestalozzi et al.}, 2004), is questioned by the
 small stellar masses implied by the rotation curves ($\la$1~\Msun) and by
 the fact that in many cases the maser spots are aligned along the direction
 of a large scale H$_2$ jet ({\it De Buizer}, 2003), suggesting that the spots could
 be part of the jet. A similar conclusion is attained from mid-IR imaging
 observations of the putative disk in NGC\,7538~IRS\,1 ({\it De Buizer and Minier},
 2005).  Nonetheless, interesting examples of convincing disks in massive
 YSOs have been found. Using SiO masers, {\it Greenhill et al.} (2004) have imaged an
 expanding and rotating disk in the Orion~KL region, while hydroxyl masers
 have been observed to trace rotation in the disk associated with the massive
 source \I\ ({\it Edris et al.}, 2005). Also, the putative disk detected towards
 Cep~A~HW2 in the \WAT\ maser, VLA study of {\it Torrelles et al.} (1998)
 has found recent confirmation in the SMA observations of {\it Patel et al.} (2005).

\bigskip
\noindent
{\em 2.2.3 Thermal lines.}
 As explained above, convincing evidence of a disk can be obtained by
 comparing its orientation to that of the associated jet/outflow.  The latter
 may be revealed by mapping the wing emission of lines such as those
 of CO and its isotopomers, HCO$^+$, SiO, and other molecular species. A disk
 tracer to be observed simultaneously with one of these outflow tracers would
 hence be of great help for disk surveys. This must be searched
 among high-excitation transitions of rare molecular species, which are the
 most suitable to trace the high density, optically thick, hot gas in the
 plane of the disk.  Experience has shown that species such as \AMM, \MCN,
 HCOOCH$_3$, \CSI\ are very effective for this purpose. In particular, the
 proximity in frequency of \MCN\ to \COI\ makes the former an ideal tool to
 study the disk {\it and} the outflow, simultaneously. With this technique,
 interferometric observations at millimeter and (thanks to the advent of the
 SMA) sub-millimeter wavelengths have been carried out, leading to the
 detection of an ever growing number of disk candidates in massive YSOs.
 A handful of sources has been observed also at centimeter wavelengths,
 with the Very Large Array: the main advantage in this regime is the possibility
 to observe strong absorption lines against the bright continuum of
 background UC \HII\ regions.
 Although successful, thermal line observations must be
 complemented by diverse imaging and spectroscopy at various wavelengths to
 safely identify a disk:  an illuminating demonstration of the effectiveness
 of this type of synergy is provided by the case of \I, which will be
 discussed in detail in Section~3.

\bigskip
\noindent
\textbf{ 2.3 Evidence for disks in high-mass YSOs}
\bigskip

In recent years, the disk searches towards massive YSOs described in the
previous section have detected an ever growing number of candidates.  The
main discriminating feature between these disks and those in pre-main
sequence low-mass stars is the ratio between disk and stellar mass. While in
the latter such a ratio is $<0.1$, for massive YSOs the disk mass becomes
comparable to or even greater than that of the star. This poses a question
about the stability and lifetime of the disks, which will be discussed in
detail in Section~4. Here, we note that candidate disks may be roughly divided
into two types: those having mass $M$ in excess of several 10~\Msun, for which
the ratio $M/M_\star\gg1$, and those with $M\la M_\star$.  As noted by {\it
Cesaroni} (2005a) and {\it Zhang} (2005), the former have in all cases
luminosities typical of ZAMS O stars, whereas the latter are more likely
associated with B-type stars.

Typically, the massive, rotating structures observed in YSOs with
luminosities $\ga10^5$~\Lsun\ have radii of 4000--30000~AU, masses of
60--500~\Msun, and observed rotation speeds of a few \kms. 
Examples are
G10.62--0.38 ({\it Keto et al.}, 1988),
G24.78+0.08 ({\it Beltr\'an et al.}, 2004, 2005),
G28.20--0.05 ({\it Sollins et al.}, 2005b),
G29.96--0.02 ({\it Olmi et al.}, 2003),
G31.41+0.31 ({\it Cesaroni et al.}, 1994; {\it Beltr\'an et al.}, 2004, 2005),
IRAS\,18566+0408 ({\it Zhang et al.}, submitted), and
NGC~7538\,S ({\it Sandell et al.}, 2003).
These objects transfer mass to the star at a rate of
$2\times10^{-3}$--$2\times10^{-2}$~\My\ ({\it Zhang}, 2005 and references
therein): for the mass range quoted above, these correspond to time scales
of $\sim10^4$~yr.
Noticeably, this is an order of magnitude less than the typical
rotation period at the outer disk radius of $\sim10^5$~yr. This fact
indicates that these massive rotating structures
cannot be centrifugally supported, and should be considered as
transient, non-equilibrium evolving structures.
These objects might be observational examples of either the
``pseudo-disks'' discussed by {\it Galli and
Shu}~(1993a, 1993b; in which however the flattening is induced by magnetic
pinching forces) or the transient structures seen in numerical
simulations of competitive accretion ({\it Bonnell and Bate}, 2005; see
also the chapter by {\it Bonnell et al.}).
On the other hand, ``true disks'', which extend
on much smaller scales, are more likely to be close to an equilibrium
state where gravity from the central star (and the disk itself) is
balanced by rotation, and, possibly, by radiative forces. Clearly, a
stability analysis makes sense only for this class of objects.  In
Section~4, we will therefore discuss the stability of massive disks,
with the aim of providing some insight into the properties of
these objects.

\begin{deluxetable}{lccccccc}
\tabletypesize{\small}
\tablecaption{List of candidate disks in high-mass (proto)stars}
\tablewidth{0pt}
\tablehead{
Name & $L(L_\odot)$ & $M(M_\odot)$ & $R$(AU) & $\Mst(M_\odot)$ & $\dot{M}_{\rm out}$(\My) & $t_{\rm out}$($10^4$~yr) & Ref.$^a$ \\
}
\startdata
AFGL\,490  & $2.2$--$4\times10^3$ &     3--6 & $\la$500 & 8--10     & $6.2\times10^{-4}$ & 0.95 & 1,2 \\
G192.16--3.82     & $3\times10^3$ &       15 &   500    & 6--10     & $5.6\times10^{-4}$ & 17 & 3,4 \\
AFGL\,5142        & $4\times10^3$ &        4 &  1800    & 12        & $1.6\times10^{-3}$ & 2 & 5,6 \\
G92.67+3.07       & $5\times10^3$ &       12 & 14400    & 4--7.5    & $1.7\times10^{-4}$ & 0.35 & 7 \\
Orion BN & $2.5\times10^3$--$10^4$ &   ?     & 500      & 7--20$^b$     & $10^{-6}$ & ? & 8,9 \\
Orion I   & $4\times10^3$--$10^5$ &    ?     & 500      & $\ga 6$   & $10^{-3}$ & ? & 10,11 \\
IRAS\,20126+4104  &    $10^4$     &        4 &  1600    & 7         & $8.1\times10^{-4}$ & 6.4 & 12,13,14,15,16 \\
G35.2--0.74N      & $10^4$        & 0.15     &     1500 & 4--7      & $3\times10^{-3}$ & 2 & 17,18,19 \\
Cep A HW2         & $\sim10^4$    & 1--8     & 400--600 & 15        & $3\times10^{-5}$ & 3 & 20,21; but see 22 \\
AFGL\,2591        & $2\times10^4$ & 0.4--1.8 & 500      & 16        & $5.8\times10^{-4}$ & 6 & 23,24,25 \\
IRAS\,18089--1732 & $6\times10^4$ &   12--45 &  1000    & $<$25     & ? & 4 & 26,27 \\
M17               &     ?     & 4--$>$110 & 7500--20000 & $<$8--20  & ? & ? & 28,29 \\
%
\enddata


\begin{flushleft}
$^a$~1: {\it Schreyer et al.} (2002, 2006);
     2: {\it Mitchell et al.} (1992);
     3: {\it Shepherd and Kurtz} (1999);
     4: {\it Shepherd et al.} (2001); \\
     5: {\it Zhang et al.} (2002);
     6: {\it Hunter et al.} (1999);
     7: {\it Bernard et al.} (1999);
     8: {\it Jiang et al.} (2005);
     9: {\it Scoville et al.} (1983); \\
    10: {\it Greenhill et al.} (2004);
    11: {\it Genzel and Stutzki} (1989);
    12: {\it Cesaroni et al.} (1997);
    13: {\it Zhang et al.} (1998b); \\
    14: {\it Cesaroni et al.} (1999a);
    15: {\it Cesaroni et al.} (2005);
    16: {\it Shepherd et al.} (2000);
    17: {\it Hutawarakorn and Cohen} (1999); \\
    18: {\it Fuller et al.} (2001);
    19: {\it Dent et al.} (1985);
    20: {\it G\'omez et al.} (1999);
    21: {\it Patel et al.} (2005); \\
    22: {\it Comito et al.} (in prep.) challenge the disk interpretation;
    23: {\it Hasegawa and Mitchell} (1995);
    24: {\it van der Tak and Menten} (2005); \\
    25: {\it van der Tak et al.} (2006);
    26: {\it Beuther et al.} (2004b);
    27: {\it Beuther et al.} (2005);
    28: {\it Chini et al.} (2004);
    29: {\it Sako et al.} (2005);

$^b$~see {\it Hillenbrand et al.} (2001) and references therein;
values depend on adopted mass-luminosity conversion.

\end{flushleft}

\end{deluxetable}

It is advisable to use a different terminology for
non-equilibrium massive rotating structures:  following {\it Cesaroni}
(2005a), we will call them ``toroids''.  What is the role of toroids in
the process of high-mass star formation? Given the large masses and
luminosities involved, they likely host a stellar cluster rather then a
single star. That is why some authors have used the term
``circumcluster'' toroids, as opposed to ``circumstellar'' disks
({\it Beltr\'an et al.}, 2005). Since their size is several times the
centrifugal radius (i.e.  the radius at which a centrifugal barrier
occurs; see, e.g., {\it Terebey et al.}, 1984), one may speculate that eventually
toroids will fragment into smaller accretion disks rotating about single
stars or binary systems ({\it Cesaroni}, 2005b).  In the following we will not
consider toroids but only ``true'' circumstellar (or circumbinary) disks in
massive YSOs.

Table~2 lists disk candidates associated with high-mass YSOs. The columns
are: luminosity, disk mass, disk radius, mass of the central star, outflow
mass loss rate, and outflow dynamical timescale.
When considering this table, several caveats are in order. Although most of
the entries are to be considered bona fide disks, in some cases the
interpretation is challenged (e.g., Cep~A~HW2; Comito et al. in prep.). Also,
some of the values given in the table are quite uncertain, in particular the
mass of the star. While in a few cases (e.g., IRAS\,20126+4104) this is
obtained directly from the Keplerian rotation curve of the disk, in the
majority of the objects the estimate comes from the luminosity or from the
Lyman continuum of the YSO. The conversion from these quantities to stellar
masses is very uncertain as it depends on the unknown evolutionary stage of
the (proto)star, on the optical depth and origin of the radio emission (\HII\
region or thermal jet), on the number of Lyman continuum photons absorbed by
dust, and on the multiplicity of the system. Hence, the luminosities are
affected by significant errors and the mass--luminosity relationship for this
type of objects is unclear. We have thus decided to adopt the values quoted
in the literature without any further analysis.

Despite the large uncertainties, one may conclude that
all objects have luminosities typical of B-type ZAMS stars, i.e. below
a few $10^4$~\Lsun. Albeit very difficult to establish, accretion
rates for disks seem to be $\sim10^{-4}$~\My\ ({\it Zhang}, 2005): this
implies a time of $\sim10^5$~yr to transfer the material from the disk to the
star, significantly larger than the rotation period ($\sim10^4$~yr), thus
allowing the disk to reach centrifugal equilibrium.  Whether this will be
stable or not is a complicated issue that we will deal with in Section~4.
Here, we just point out that the disk time scale is comparable to the typical
free-fall time of molecular clumps hosting massive star formation, which in
turn equals the star formation time scale predicted by recent theoretical
models ($\sim10^5$~yr; {\it Tan and McKee}, 2004). This seems to suggest
that the material lost by the disk through infall onto the star and ejection
in the jet/outflow, is continuously replaced by fresh gas accreted from the
surrounding envelope.

Na\"{\i}vely, one may thus conclude that disks fit very well in the high-mass
star formation scenario. However, this seems to hold only for B-type stars,
as the stellar masses quoted in Table~2 are never in excess of $\sim$20~\Msun
and the luminosities on the order of $\sim10^4$~\Lsun, typical of stars
later than B0. What about disks in more luminous/massive objects?
Disk searches performed in sources with luminosities typical of
O-type stars (i.e. above $\sim10^5$~\Lsun) resulted in negative detections:
in all cases, as already discussed, massive rotating toroids were found. 
One possibility is that the latter ``hide'' true disks in their interiors,
still not detectable due to insufficient angular resolution and sensitivity.
It must be taken into account that objects that luminous are mostly located
at significantly larger distances than B stars. A noticeable exception is
Orion, the closest high-mass star forming region (450~pc), whose
total luminosity is $\sim10^5$~\Lsun. The presence of a disk rotating about
source I has been
clearly established by VLBI observations of the SiO masers (see Table~2).
However, it remains unclear which fraction of the total luminosity is to
be attributed to this object. Thanks to their recent high-frequency
observations with the SMA, {\it Beuther et al.} (2006) have interpreted the
continuum spectrum of source I as the combination of free-free emission from
a uniform \HII\ region plus thermal emission from dust. According to their
fit, one expects an optically thin flux of $\sim$45~mJy at $\sim$200~GHz,
corresponding to a B1 ZAMS star. On the other hand, proper motion
measurements of source I and BN ({\it Rodr\'{\i}guez}, 2005) have demonstrated that
the two are receding from each other with speeds respectively of 12 and
27~\kms. Since the estimated mass of BN lies in the range
$\sim$7--20~\Msun\ (see Table~2), the ratio between the two velocities
suggests a mass for source I of 16--45~\Msun. While the upper limit is larger
than expected for a ZAMS star of $\sim10^5$~\Lsun, the lower seems roughly
consistent with the estimate from the free-free emission. It is hence
possible that Orion~I is an early B or late O star. In this case, we would
be still missing an example of a disk in an early O-type star.

Is such a lack of evidence for disks in O-type stars due to an observational
bias? The absence of disks in O stars could have important theoretical
implications, as disks are believed to play a key role in favouring accretion
onto the star through angular momentum dissipation.  If no disks are present
in stars more massive than $\sim$20~\Msun, alternative models of massive star
formation may be needed, such as, e.g., those involving coalescence of lower
mass stars ({\it Bonnell and Bate}, 2005). It is hence in order to discuss
all possible technical limitations which may have hindered detection of disks
in O-type stars.

\bigskip
\noindent
\textbf{ 2.4 Are disks in O-type stars detectable?}
\bigskip

As noted above, no clear evidence for circumstellar disk in early O-type stars
has been presented to date. But even disks in B-type stars do not seem to be
easy to detect. It is difficult to draw any statistically reliable
conclusion on the real number of high-mass YSOs with disks, as most studies
have been conducted towards selected candidates. Only one systematic search
for disks is known to date: this is the recent survey by {\it Zhang et al.} (pers.
comm.). These authors used the VLA to survey 50 YSOs with luminosities
between $10^3$ to $10^5$~\Lsun\ in \AMM\ inversion transitions, and detected
10 possible disks. Is a detection rate of 20\% the one expected if all B-type
stars form through disk accretion? And are disks in O-type stars really
elusive?

We consider the possibility of
an observational bias in disk searches, due to limited sensitivity,
angular, and spectral resolution (in the case of line observations).
Let us discuss separately the sensitivity and resolution issues.

\bigskip
\noindent
{\em 2.4.1 Spectral and angular resolution.}
As explained in Section 2.2, most disk searches attempt to detect the velocity
gradient due to rotation about the star. To achieve such a detection, the
observations must disentangle the emission of the red- and blue-shifted parts
of the disk, not only in space but also in velocity. One
constraint is hence the angular resolution achieved by current
interferometers. As for the spectral resolution, at radio wavelengths one may easily
attain 0.1~\kms, sufficient to resolve lines as broad
as a few \kms.  The limitation is due to the line width, of
the same order as the expected rotation velocity.
Therefore, to detect rotation with velocity $V$ at radius $R$,
two conditions must be fulfilled: the separation between two diametrically
opposite points must be greater than the
instrumental half power beam width (HPBW); and the corresponding velocity
difference must be greater than the
line full width at half maximum (FWHM). These may be expressed as:
\begin{eqnarray}
2\,R & > & \Theta\,d \\
2\,V(R)\,\sin\,i & > & W    \label{evel}
\end{eqnarray}
where $i$ is the inclination angle with respect to the line of sight ($i=0$ for
a face-on disk), $\Theta$ the HPBW, $d$ the distance to the source, and $W$
the line FWHM. For a star with mass $\Mst$, $V=\sqrt{G\,\Mst/R}$, so that
one obtains
\begin{equation}
d({\rm kpc}) < 7 \frac{\Mst(\Msun)\,\sin^2i}{\Theta(\arcsec)\,W^2(\kms)}.
		      \label{edis}
\end{equation}
A conservative estimate of this expression may be obtained for
$\Theta=1\arcsec$ (the HPBW of millimeter interferometer), $W=5$~\kms (the
typical line FWHM of a hot molecular core -- to be taken as an upper limit,
as the intrinsic line width is less than the observed one), and a mean value
of $\langle\sin^2i\rangle=2/3$ assuming random orientation for the disk
axes.  The result is $d({\rm kpc}) < 0.19 \Mst(\Msun)$ which implies that the
maximum distance at which a disk can be detected in, e.g., a 50~\Msun\ star
is 9.5~kpc. This shows that disks in all massive stars should be seen up to
the Galactic center.

The previous analysis does not take properly into account the effect of the
inclination angle, as well as other effects which may complicate the
picture: infall and outflow, disk flaring, and non-Keplerian
rotation. Further complication is added by the presence of multiple
(proto)stars, each of these possibly associated with and outflow/disk,
and by the fact that molecular species believed to be ``pure'' disk tracers
might instead exist also in the outflow.

Na\"{\i}vely, infall and rotation speeds should both be $\propto R^{-0.5}$ and
hence comparable, so that the Keplerian pattern should be only slightly
affected. However, recent observations have revealed infall with no or little
rotation, in the O-type (proto)stars, G10.62--0.38 ({\it Sollins et al.}, 2005a) and
IRAS\,18566+0408 ({\it Zhang} pers. comm.). This occurs on spatial scales of
$\sim$1000~AU, comparable to those over which pure rotation is seen in the
B-type sources of Table~2.  Possible explanations must involve
angular momentum dissipation, perhaps due to magnetic field braking.

Deviation from Keplerian rotation is expected to occur at large radii, where
the enclosed disk mass becomes comparable to the stellar mass (see the
example in Section~3): in this case, though, the value of $V$ is larger
than in the Keplerian case and Eq.~(\ref{evel}) is satisfied.

Outflows could ``spoil'' the detection of disks, because the ratio between
expansion and rotation speeds may be as large as $\sim$100. However, the
molecular lines used for disk searches do not seem to trace also the
outflow/jet: an illuminating example of this is given by the best studied
disk-outflow/jet system to date, \I: here the Keplerian disk (rotating at a
few \kms; {\it Cesaroni et al.}, 2005) is clearly seen in the \MCN\ lines,
notwithstanding the presence of a jet with velocities in excess of
100~\kms\ ({\it Moscadelli et al.}, 2005) on the same scale as the disk. It
is hence clear that the success of disk searches and the possibility to
decouple them from the associated outflows/jets depend crucially on the
molecular tracers used.

Finally, disk flaring may help revealing rotation. In fact, while
Eq.~(\ref{edis}) depends sensitively on the inclination angle $i$, this
result is obtained for an infinitesimally thin disk.  Real
disks have a finite $H(R)$ and, considering as an example IRAS\,20126+4104,
we can assume $H=R/2$ (see Section~4.2).  Under
this assumption, the line of sight will lie in the plane of the disk as long
as $76^\circ<i<104^\circ$.  For a random orientation of the disk axis, this
implies that $\sim$24\% of the disks will satisfy this condition.
Noticeably, such a number is very close to the 20\% detection rate obtained
in the previously mentioned \AMM\ survey by {\it Zhang et al.} and suggests that
disk inclination should not represent a serious limitation for disk
searches.

\bigskip
\noindent
{\em 2.4.2 Sensitivity.}
The continuum emission from disks in the (sub)millimeter regime (see also
Section~2.2) should be easily detected by modern interferometers. This can be
evaluated assuming a disk mass $M$ proportional to the stellar
mass $M_\star$, e.g., $M=M_\star/2$ (see Table~2) and a dust
temperature of 100~K.  The flux density measured at a distance $d$ is
equal to
\begin{equation}
S_\nu({\rm mJy}) = 85.3 \, 
		   \left[\frac{\nu({\rm GHz})}{230.6}\right]^{2+\beta}
		   \, \Mst(\Msun) \, d^{-2}({\rm kpc})
\end{equation}
where a dust absorption coefficient equal to
0.005~cm$^2$\,g$^{-1}$~$[\nu({\rm GHz})/230.6]^\beta$
has been adopted ({\it Kramer et al.}, 1998). Very conservatively, one may
assume a sensitivity of 30~mJy at 230.6~GHz, which sets the constraint
$d({\rm kpc}) < 1.7 \sqrt{\Mst(\Msun)}$. Disks associated with stars of 10
and 50~\Msun\ should be detectable up to 5.4 and 12~kpc, respectively.
Therefore, the main limitation is not given by the sensitivity but by
confusion with the circumstellar envelope, as illustrated in Section~2.2.

On the other hand, line observations are less affected by this
problem, as the velocity information helps decoupling the rotating disk from
the more quiescent envelope. The line intensity may be estimated under the
assumption of optically thick emission. For $T\ga100$~K, the
flux density from the disk surface within a radius $R$ must be at least
\begin{equation}
S_\nu = \frac{R^2 \, \pi B_\nu(T)}{4 \, \pi \, d^2}    \label{elin}
\end{equation}
with $B_\nu$ black-body brightness. Here, we have assumed $R^2$ as a lower
limit to the surface: this relies upon the fact that the disk thickness is
$\sim R/2$, so that the effective surface ranges from $R^2$ (edge on) to $\pi
R^2$ (face on). In order to detect the line emission from the disk with an
instrumental noise $\sigma$, one must have $S_\nu>3\,\sigma$, which using
Eq.~(\ref{elin}) turns into $R > \sqrt{12 \sigma/B_\nu(T)} d$. The radius within
which the flux is measured must satisfy also Eq.~(\ref{evel}). For the
sake of simplicity, we assume a temperature $T=100$~K independent of $R$
and a noise $\sigma\simeq0.1$~Jy at 230~GHz, thus obtaining
\begin{equation}
d({\rm kpc}) < 6.2 \frac{\Mst(\Msun) \sin^{2}i}{W^2(\kms)}
\end{equation}
For the same fiducial values adopted for Eq.~(\ref{edis}), this condition
takes the form $d({\rm kpc})<0.16\,\Mst(\Msun)$. Once more, all stars of a few
10~\Msun\ should be detectable up to the Galactic center.

In conclusion, albeit very rough, the previous discussion suggests that the
currently available instruments should be sufficient to detect circumstellar
disks around all massive stars, if the disk mass is non-negligible with
respect to the stellar mass. One must hence look for an astronomical
explanation to justify the lack of detections for disks in O-type stars.
As illustrated in the following, this might be found in the
stability and/or lifetime of massive disks.

\bigskip

\centerline{\textbf{ 3. \I: THE PROTOTYPE DISK}}
\bigskip

Among the disk candidates listed in Table~2, one stands unique as the best
studied and probably most convincing example of a Keplerian disk rotating
about a massive YSO. This is \I, an IRAS point source believed to be
associated with the Cyg~X region ({\it Wilking et al.}, 1989 and references
therein).  First recognized by {\it Cohen et al.} (1988) as an OH maser
emitter, it was then observed in the continuum, at millimeter wavelengths
({\it Wilking et al.}, 1989; {\it Walker et al.}, 1990), and in the CO(2--1)
line by {\it Wilking et al.} (1990), who detected a powerful, parsec-scale
molecular outflow. Later on, association with \WAT\ and \METH\ masers was
also established ({\it Palla et al.}, 1991; {\it MacLeod and Gaylard}, 1992)
and the molecular emission from high-density tracers was detected ({\it
Estalella et al.}, 1993). These pioneering works provided evidence that the
source is associated with a luminous, embedded star still in a very early
stage of the evolution. The fact that no free-free continuum was detected
({\it Tofani et al.}, 1995) indicated that although luminous, the source had
not yet developed an \HII\ region, while the presence of a molecular outflow
suggested that the YSO could be actively accreting material from the parental
cloud. Eventually, observations at 3~mm with the Plateau de Bure
interferometer ({\it Cesaroni et al.}, 1997) achieved the angular resolution
needed to dissect the outflow and analyse the velocity field in the
associated molecular core. Thanks to these observations, it was first
recognized that the core was rotating about the outflow axis.

\begin{figure}
 \epsscale{1.0}
 \plotone{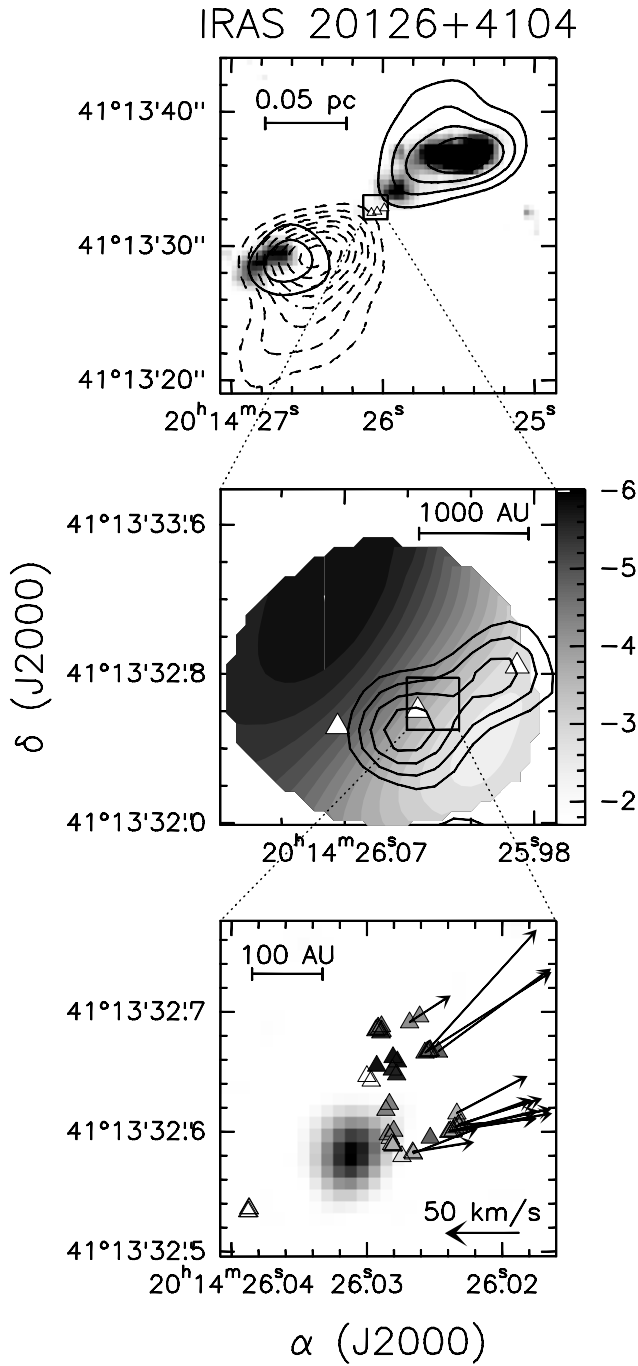}
 \caption{\small 
 Disk/outflow system in the high-mass (proto)star IRAS\,20126+4104.  Top:
 Overlay of the H$_2$ lime emission at 2.2~$\mu$m (grey scale) and the bipolar
 outflow traced by the HCO$^+$(1--0) line ({\it Cesaroni et al.}, 1997). Solid and
 dashed contours correspond respectively to blue- and red-shifted gas.
 The triangles
 mark the positions of the H$_2$O maser spots detected by {\it Tofani et al.}
 (1995).  Middle: 3.6~cm continuum map (contours; {\it Hofner et al.}, 1999)
 overlaied on a map of the velocity measured in the C$^{34}$S(5--4) line by
 {\it Cesaroni et al.} (2005).  Bottom: Distribution of the H$_2$O maser spots
 ({\it Moscadelli et al.}, 2000, 2005) compared to a VLA map (image) of the 7~mm
 continuum emission ({\it Hofner}, pers. comm.). The grey scale of the spots ranges
 from white, for the most red-shifted spots, to black, for the most
 blue-shifted. The arrows denote the absolute proper motions of the spots,
 measured by {\it Moscadelli et al.} (2005).
 }
 \label{f20126}
\end{figure}

This finding triggered a burst of observations towards \I, thus shedding
light on its properties and allowing us to draw a detailed picture which is
best summarized in Fig.~\ref{f20126}.  The main features of this object may
be summarized as follows.

{\bf (1)}~~ The spectral energy distribution, well sampled at (sub)millimeter
 and mid-IR wavelengths ({\it Cesaroni et al.}, 1999a and references therein)
 supplies us with a luminosity of $\sim10^4$~\Lsun\ for a distance
 of 1.7~kpc. Although near-IR observations of the 2.2~$\mu$m continuum reveal
 an embedded cluster spread over 0.5~pc ({\it Cesaroni et al.}, 1997), arcsecond
 resolution mid-IR images ({\it Cesaroni et al.}, 1999a; {\it Shepherd et al.}, 2000;
 {\it Sridharan et al.}, 2005) prove that most of the luminosity arises from the inner
 1000~AU.

{\bf (2)}~~ A bipolar molecular outflow is seen in various tracers, including
 CO and HCO$^+$ ({\it Wilking et al.}, 1990; {\it Cesaroni et al.}, 1997, 1999b;
 {\it Shepherd et al.}, 2000). The kinematical outflow age
 is $t_{\rm out}\simeq 6\times10^4$~yr, which may be taken as a lower
 limit to the age of the YSO powering the flow. The momentum
 (400~\Msun\,\kms) and mass loss rate ($8\times10^{-4}$~\My), as well
 as the other parameters of the flow, are typical of outflows in
 high-mass YSOs.

{\bf (3)}~~ The orientation of the outflow axis changes by $\sim45^\circ$
 from the large (1~pc) to the small (0.1~pc) scale. A similar behaviour is
 observed in H$_2$ knots which are distributed along the outflow axis and
 thus describe an S-shaped pattern centred on the YSO powering the flow ({\it
 Shepherd et al.}, 2000). These features supply evidence that one is
 observing a precessing jet feeding the outflow, the former outlined by the
 shocked H$_2$ emission, the latter by the CO and HCO$^+$ line emission. {\it
 Cesaroni et al.} (2005) estimate a precession period of $2\times10^4$~yr,
 which implies that the jet/outflow has undergone at least 3 full
 precessions.

{\bf (4)}~~ The collimated jet traced by the H$_2$ knots is also seen on the same
 scale in other molecular tracers such as SiO ({\it Cesaroni et al.}, 1999a; {\it Liu et
 al.}, 2005), \AMM\ ({\it Zhang et al.}, 1999) and \METH\ ({\it Cesaroni et al.}, 2005),
 although the morphology and kinematics of the gas may vary significantly
 from molecules tracing the pre-shock and those arising from the post-shock
 material ({\it Cesaroni et al.}, 2005). The expansion speed of the jet over 0.5~pc
 is estimated $\sim$100~\kms\ ({\it Cesaroni et al.}, 1999a), while the component
 along the line of sight is relatively small, of order $\pm20$~\kms. This
 finding and the fact that both blue- and red-shifted emission overlap in
 space prove that the jet/outflow axis inside $\le0.5$~pc from the YSO lies
 very close to the plane of the sky, at an angle $\la9^\circ$.  Proper motion
 measurements of the \WAT\ masers indicate a jet expansion speed in excess of
 $\sim100$~\kms\ at a distance of $\sim250$~AU from the powering source
 ({\it Moscadelli et al.}, 2005), while on an intermediate scale between the one
 traced by H$_2$ (and other molecular lines) and that sampled by the
 \WAT\ masers, free-free emission from the thermal component of the jet is
 seen at 3.6~cm ({\it Hofner et al.}, 1999). All these results indicate that the
 jet/outflow system extends almost continuously from the
 neighbourhoods of the YSO powering it, to the outer borders of the parental
 molecular clump.

{\bf (5)}~~ At the geometrical center of the bipolar jet/outflow, a hot molecular
 core is detected, of $\sim200$~K. When observed at high angular resolution
 in the \MCN\ transitions, this core presents a velocity gradient roughly
 perpendicular to the jet. {\it Cesaroni et al.} (1997) interpreted this result as
 rotation about the YSO powering the outflow, while subsequent imaging in
 \MCN(12--11) ({\it Cesaroni et al.}, 1999a), \AMM(1,1) and (2,2) ({\it Zhang et al.},
 1998b) and \CSI(2--1) and (5--4) ({\it Cesaroni et al.}, 2005) has established that
 the rotation is Keplerian. However, the estimate of the stellar mass seems
 to differ depending on the tracer used to measure the rotation curve. In
 fact, going from a few $10^3$~AU to $10^4$~AU, such an estimate changes from
 7~\Msun\ ({\it Cesaroni et al.}, 2005) to 24~\Msun\ ({\it Zhang et al.}, 1998b). This
 effect may be due to the disk mass enclosed inside the radius at which the
 velocity is measured ({\it Bertin and Lodato}, 1999): for small radii, the disk
 mass inside that radius ($\sim4$~\Msun) is less than that of the star,
 whereas at large radii the two become comparable, thus mimicking Keplerian
 rotation about a bigger star. {\it Cesaroni et al.} (2005) have also estimated the
 temperature profile as a function of radius, which seems to be compatible
 with the ``classical'' law expected for geometrically thin disks heated
 externally by the star or internally by viscosity: $T\propto R^{-3/4}$. The
 mean disk temperature is of order $\sim170$~K.  Additional evidence of a
 circumstellar or possibly circumbinary disk in \I\ comes also from OH and
 \METH\ maser studies ({\it Edris et al.}, 2005), as well as from recent near and
 mid-IR continuum images obtained by {\it Sridharan et al.} (2005). In
 particular, at 2.2~$\mu$m the disk is seen in absorption as a dark
 silhouette similar to those observed in the optical towards the proplyds in
 the Orion nebula (e.g., {\it O'Dell and Wen}, 1994).

In conclusion, the observational results obtained for \I\ provide robust
evidence for the existence of a Keplerian disk associated with a precessing
jet/outflow powered by a massive YSO of $\sim10^4$~\Lsun. One possibility
is that one is dealing with a binary system, as suggested by the jet
precession which may be caused by interaction with a companion. Indeed, the
latter seems to appear as a secondary peak, close to the disk border, in the
mid-IR images by {\it Sridharan et al.} (2005). However, it is unlikely that the
mass of the companion is comparable to that of the YSO at the center of the
disk, because the Keplerian pattern does not seem to be significantly
perturbed by the presence of the second star. Better angular resolution and
sensitivity are required to settle the binary issue.

\bigskip

\centerline{\textbf{ 4. FORMATION AND STABILITY OF MASSIVE DISKS}}
\bigskip

In this section, we consider some theoretical
problems posed by the existence of massive  (few solar masses) disks around
massive  protostars as in the cases listed in Table~2. One obvious question
is that of their stability particularly if, as seems likely, the disk mass in
some cases is comparable to the stellar mass. A second problem
is posed by the high
accretion rates needed both to form the massive stars on a reasonable
timescale (typically $10^5$ years, according to {\it Tan and McKee}, 2004)
and to account for the observed large outflow rates. We ask the question of
whether classical ``$\alpha$-disk models'' can account for such high
accretion rates and conclude it is unlikely.  Thirdly, following the approach
of {\it Hollenbach et al.} (2000), we consider briefly effects which might limit
massive disk lifetimes in the  phase just subsequent to the cessation of
accretion. We commence however by a brief discussion of what one might
naively expect to be the properties of disks associated with massive
protostars.

\bigskip
\noindent
\textbf{ 4.1 From clouds to disks}
\bigskip

 From a theoretical point of view, the presence of disks
around massive protostars is expected on the same physical grounds as
in the case of solar mass protostars. In a simple inside-out collapse
with constant accretion rate ({\it Terebey et al.},~1984), the mass of the
disk increases linearly with time, $M\propto t$, whereas the disk
radius increases much faster, $R\propto t^3$, or $R\propto M^3$. 
Thus, in principle, we expect massive, centrifugally supported disks to
have much larger sizes than the low-mass disks observed around T Tauri
stars.

In particular, the sizes and masses of the disks observed around young
massive stars can be used to set constraints on the physical
characteristics of the clouds from which they formed. It is a good
approximation to express the disk angular velocity at a radius $R$ as
$\Omega (R) \approx (GM_{\rm t}/R^3)^{1/2}$ ({\it Mestel}, 1963), where $M_{\rm t}$ is the
total (star plus disk) mass of the system (the relation holds exactly
only for a surface density $\Sigma\propto R^{-1}$).
The specific angular momentum of the system is then
$J/M_{\rm t}\approx \Omega R^2 \approx (GM_{\rm t} R)^{1/2} \approx
10^{-2}$~km~s$^{-1}$~pc with the values listed in Table~2 for \I.
This is within the range of observed values of the specific angular
momentum of molecular cloud cores in low-mass star forming regions
($J_{\rm c}/M_{\rm c}\approx 10^{-3}$--$10^{-1}$~km~s$^{-1}$~pc;
{\it Goodman et al.},~1993), although it is unclear if similar
values pertain to massive star-forming regions too.

In any case, the formation of disks like \I\ from the collapse of
slowly rotating clumps appears physically possible, and seems to imply
that the clump's angular momentum is not much reduced during the
formation of a massive disk. It is possible to obtain the radius
$R_{\rm i}$ within which the mass $M_{\rm t}$ of the star plus disk
system was originally contained in the parental cloud, $R_{\rm
i}\gtrsim (G M_{\rm t} R/\Omega_{\rm c}^2)^{1/4}$, where $\Omega_{\rm
c}$ is the cloud's angular rotation. This imposes a severe constraint
on the effective sound speed in the cloud $c_{\rm eff}\approx GM_{\rm
t}/2R_{\rm i} \lesssim (G^3 M_{\rm t}^3 \Omega_{\rm
c}^2/2R)^{1/8}\approx 1$--2~km~s$^{-1}$ for $\Omega_{\rm c}\approx
1$~km~s$^{-1}$~pc$^{-1}$.  Notice that the resulting accretion
rate $\dot M\approx c_{\rm eff}^3/G$ is $\dot M\approx 2\times
10^{-4}$~\Msun~yr$^{-1}$, which implies a timescale of $\sim 6\times
10^4$~yr for the formation of the \I\ star plus disk system, in
agreement with the inferred outflow kinematical age (see Table~2).
Thus, the cloud cores where massive stars are formed must be
characterized, at least in their central parts, by low or moderate
levels of turbulence, similarly to their low-mass counterparts. The
cold, massive ($M_{\rm c}\approx 10^2$~\Msun) molecular cloud cores
with narrow linewidhts ($\Delta v\approx 1$~km~s$^{-1}$) recently
discovered by {\it Birkmann et al.}~(2006) seem to present the required
properties, and suggest that the initial conditions for the formation
of massive stars should be similar to those observed in low-mass star
forming regions.

Numerical calculations of the collapse of massive molecular
cores have been performed by {\it Yorke and Sonnhalter}~(2002),
who also included a detailed treatment of continuum radiation transfer. The
calculations start from cold, massive, slowly rotating clouds
far from equilibrium that typically contain tens of Jeans masses.  The
resulting mass accretion rate is strongly time-dependent, peaking at a
value $\dot M\approx 10^{-3}$~\Msun~yr$^{-1}$ after $\sim 10^4$~yr from
the onset of collapse and declining thereafter because of radiation
pressure.  In some cases, the infalling material flows onto a disklike
feature appearing after $\sim 10^5$~yr from the onset of collapse, which
rapidly extends up to $~10^4$~AU in radius. The calculated density in
the outer disk is $\sim 10^{-18}$~g~cm$^{-3}$ and compares well with
that inferred from the ratio of C$^{34}$S (2--1) and (5--4)
lines measured in \I\ at radii $R\gtrsim 2000$~AU.  The
calculations suggest that for disks around stars above $\sim
20$~\Msun\ stellar radiative forces may contribute as much as rotation
to the radial support of the disk.  The disk may also be short-lived
disappearing after some $10^4$~yr.

\bigskip
\noindent
\textbf{ 4.2 Local gravitational stability of massive disks}
\bigskip

It is well known, since {\it Toomre} (1964), that a fluid, thin disk is
unstable to axisymmetric gravitational disturbances if the stability
parameter 
\begin{equation}
Q=\frac{c_s\kappa}{\pi G \Sigma}<1,
\end{equation}
where $c_s$ is the sound speed, $\Sigma$ the disk surface
density, and $\kappa$ the epicyclic frequency. The angular velocity and the
epicyclic frequency are related by $\kappa=f\Omega$, where $f$ is a numerical
factor dependent on the shape of the rotation curve: for Keplerian rotation
$f=1$, while for a flat rotation curve $f=\sqrt{2}$

The value of $Q$, the disk aspect ratio and the total disk mass can be
easily related to one another in the following way. The disk thickness
$H$ is given by
\begin{equation}
H=\left\{
\begin{array}{ll}
c_s/\Omega & \mbox{if $M/M_{\rm t} \ll H/R$} \\
c_s^2/\pi G\Sigma & \mbox{if $M/M_{\rm t}\gtrsim H/R$}.
\end{array}
\right.
\end{equation}
If we assume a power-law behaviour for $\Sigma\propto R^{-1}$, so
that $M(R)=2\pi\Sigma R^2$, and we adopt the approximation
$\Omega\approx (GM_{\rm t}/R^3)^{1/2}$, we can rewrite $Q$ as
\begin{equation}
Q = \left\{
\begin{array}{ll} 
\frac{2H}{R}\frac{M_{\rm t}}{M} \gg 1 & 
\mbox{if $\frac{M}{M_{\rm t}} \ll \frac{H}{R}$} \\
\left(\frac{2H}{R}\frac{M_{\rm t}}{M}\right)^{1/2} \lesssim \sqrt{2} & 
\mbox{if $\frac{H}{R} \lesssim \frac{M}{M_{\rm t}} \ll 1$} \\
\left(\frac{4H}{R}\frac{M_{\rm t}}{M}\right)^{1/2} \simeq 2\sqrt{\frac{H}{R}} & 
\mbox{if $\frac{M}{M_{\rm t}}\approx 1$}
\end{array}
\right.
\end{equation}

The relationship between cumulative disk mass, aspect ratio and
stability parameter $Q$ is also shown in Fig. 2, that shows a contour
plot of $Q$, as a function of $M/M_{\rm t}$ and of $H/R$ based on the
equation above. We wish to stress that $Q$ is only a measure of the
local stability of a massive disk: the above relationship and the
figure describe the stability of the disk at a radius $R$, where
$M$ is the disk mass enclosed within $R$ and the aspect ratio
$H/R$ is computed at $R$.  We then see that, in order for the disk to
be gravitationally unstable, we need $M/M_{\rm t}\gtrsim 2\,H/R$.

\begin{figure}[hbt!]
\epsscale{1.0}
\plotone{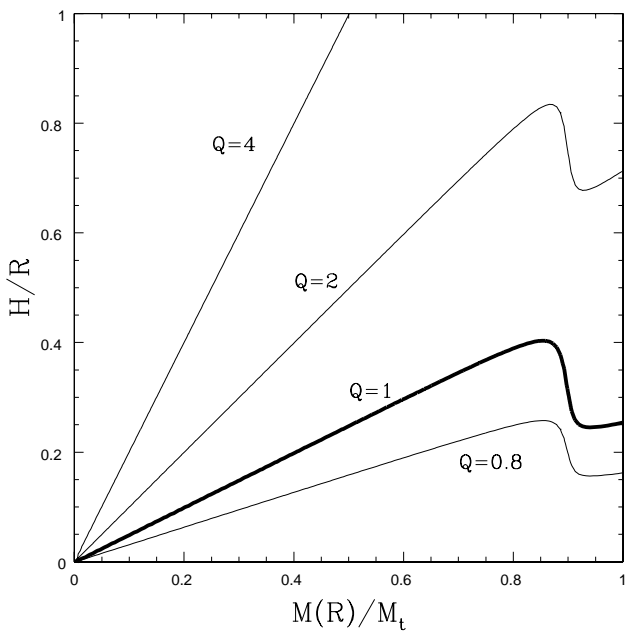}
\caption{\small 
Contour plot of the relationship between stability parameter $Q$,
enclosed disk mass $M(R)/M_{\rm t}$ and aspect ratio $H/R$.
This plot describes local stability at radius $R$, and $M$ has
to be intended as the mass enclosed within $R$. The curves are computed
under the assumption that $\Sigma \propto R^{-1}$. The wiggles
at large disk mass indicate the transition from a Keplerian to a flat
rotation curve. The thick contour corresponds to $Q=1$, which marks the
threshold between stability and instability.}
\label{Qplot}
\end{figure}

Non-axisymmetric, spiral disturbances are generally more unstable, and
thin disks can result unstable at considerably larger values of
$Q\approx 3$--4. On the other hand, it is also well known that a finite
thickness is a strong stabilizing effect. The general problem of
finding the marginal stability of a thick disk with respect to
non-axisymmetric instabilities is however too complex to be treated
analytically. Numerical simulations ({\it Lodato and Rice}, 2004, see below)
seem to show that relatively thick and massive disks evolve in such a
way to achieve $Q\approx 1$. 

The simple estimate of $Q$ outlined above and shown in Fig. 2
offers an easy tool to evaluate the local stability of the massive disks
observed around massive stars. Let us take, as an
example, the case of \I\ and consider its properties at a radius
$R\approx 1600$ AU. As we have seen before, the enclosed mass within
1600 AU is of the order of 4~\Msun, while the central object mass is
7~\Msun, so that $M(R)/M_{\rm t}\approx 0.4$.  The average temperature
at $R$ is 170~K, so that $H/R\approx 0.4$. With these estimates, the
value of $Q$ turns out to be $Q\approx 2$ (see Fig. 2). The disk is
therefore expected to be stable with respect to axisymmetric disturbances
but unstable with respect to spiral instabilities.

\bigskip
\noindent
\textbf{ 4.3 Transport properties and evolution of massive disks}
\vspace*{1mm}

 From theoretical models of massive star formation ({\it Yorke and Sonnhalter},
2002) we know (see above) that in order to form massive stars, the disk
has to be fed at high accretion rates (of the order of
$10^{-3}$~\My). On the other hand, we also know from
observations that this mass has to be transported at similar rates to
small radii, in order to resupply the powerful outflows observed in
these systems (with outflow rates in the range of
$10^{-4}-10^{-3}$~\My, see Table 2). The natural question
that arises is therefore whether in the disk there are efficient
mechanisms able to deliver the torques needed to redistribute the
angular momentum within the disk at the required rates.

Torques in accretion disks are generally parametrized through the
dimensionless parameter $\alpha$, that measures the strength of the
viscous torques relative to the local disk pressure. The most promising
mechanism of transport in disks is usually considered to be related to
MHD instabilities, and in particular to the magneto-rotational
instability ({\it Balbus and Hawley}, 1992). This kind of instability is able
to provide torques with $\alpha\approx 10^{-2}$ (see {\it Balbus}, 2003 for a
review). It is not clear, however, how efficient such processes can be
in the cold outer regions of the disk, where the ionization level is
expected to be small.

On the other hand, we have just shown how the massive disks observed
around massive stars are likely to be gravitationally unstable. This
leads to another important source of angular momentum transport, in the
form of gravitational instabilities. 
In this context, numerical
simulations play a very important role, being the only way to follow
the dynamics of gravitationally unstable disks to the non-linear
regime.
A detailed discussion of the
role of gravitational instabilities is presented in the chapter by
{\it Boss et al.}.
Here, we will only summarize the main results on this
issue, in consideration of the particular properties of disks around
massive stars (taking, as a prototypical example, the case of \I).

{\it Laughlin and Bodenheimer} (1994), in a pioneering work, have shown for
the first time through smoothed particle hydrodynamics simulations the
effectiveness of gravitational instabilities in promoting the accretion
process. They found that the redistribution of matter in the disk, due
to the spiral structure, occurs over a few rotational periods.
However, these early simulations were limited by the particular choice
of the equation of state of the gas in the disk. {\it Laughling and
Bodenheimer} (1994), in fact, assumed that the disk is isothermal, thus
inhibiting the important feedback effect on the disk stability provided
by the heating of the disk due to the instability itself.

A full treatment of the disk thermodynamics, including a detailed
radiation transfer through the disk, is presently impossible to
achieve with current numerical techniques (although progress is being
made to include a more realistic cooling, see {\it Johnson and Gammie},
2003).  The approach that has been taken more recently is to adopt some
simplified prescription for the cooling of the disk and
constrain the evolution under such simplified conditions
({\it Laughlin and Korchagin}, 1996; {\it Pickett et al.}, 2000; {\it Gammie}, 2001;
{\it Lodato and Rice}, 2004, 2005).  {\it Gammie} (2001) assumed that $t_{\rm
cool}=\beta\Omega^ {-1}$ and found that for $\beta<3$ the disk
fragments into bound objects, whereas for $\beta>3$ a quasi-steady
unstable state is reached, in which efficient redistribution of
angular momentum takes place. {\it Lodato and Rice} (2004, 2005) have extended
the analysis of Gammie to a full 3-dimensional, global context. In this
way they were able to study the effect of the development of global
spiral structures on the disk evolution, which is precluded in the
local simulations by Gammie. The general result of these simulations
(either local or global) is that the gravitational instability
saturates at an amplitude such that the dissipation provided by the
instability is able to balance the imposed cooling. The typical values
of $\alpha$ found in these simulations are in the range $0.01-0.06$.
More recently, {\it Rice et al.} (2005) have elucidated the
process of fragmentation in massive disks, finding that a
self-gravitating disk can provide a stress no larger than
$\alpha\approx 0.06$. If the cooling time is so short that the
dissipation required to balance it is larger than this maximum, the
response of the disk is to fragment.

We can now estimate the maximum accretion rate expected for a  
centrifugally supported accretion disk. Assuming a density profile
$\Sigma\propto R^{-p}$, we have
\begin{equation}
\dot{M}=3\pi\nu\Sigma\approx {\alpha \over 2-p}
\left(\frac{H}{R}\right)^2 M(R)\Omega(R),
\label{accr}
\end{equation}
where $\nu=\alpha H^2\Omega$ is the viscosity, expressed through the
{\it Shakura-Sunyaev} (1973) prescription, and $M(R)$ is the disk mass
enclosed within $R$.  Let us consider, as an illustration, the case of
\I, where the mass enclosed within $R=1600$ AU is $M=4$~\Msun, and the
aspect ratio is $H/R \approx 0.4$ (see Section~4.2).
With $\alpha=0.06$ and $p=1$
we find that the maximum accretion rate that can be delivered by
gravitational instabilities is $\dot{M}\approx
10^{-5}$~\Msun~yr$^{-1}$, much smaller than the values required to
power the outflow (the MRI instability can only provide lower accretion
rates). It is important to stress that the above numerical
estimate for the mass accretion rate depends on the specific radius at
which we evaluate the enclosed mass and the angular velocity at that
radius. The ``problem'' that we are referring to here refers to the
difficulty of transferring to the star $\sim 4$~\Msun\ from a distance of
$\sim 1600$~AU through a standard accretion disk.  The problem may be
alleviated (or eliminated) if the density distribution in the disk is
very steep ($p\approx 2$) and most of the disk mass is concentrated at
small radii where the rotation period is shorter and therefore the
disk's dynamical evolution is faster.

The required high accretion rates pose in general a serious challenge
for theoretical models. However, we must keep in mind that the
accretion rates predicted by hydrodynamical or magnetohydrodynamical
disk simulations is highly variable in time, and may be temporarily
enhanced by an order of magnitude above the average value.  For
example, {\it Lodato and Rice} (2004, 2005) have compared simulations of
self-gravitating disks of low ($M \ll M_\star$) and high mass
($M\approx M_\star$), finding that low-and high-mass disks
behave somewhat differently. In the low-mass case a self-regulated
state is rapidly achieved and the evolution of the disk after the
development of the gravitational instability is quasi-steady. On the
other hand, as $M$ approaches $M_{\star}$ the temporal
behaviour of the disk becomes more complex and a series of recurrent
episodes of spiral activity are often seen, lasting for roughly one
rotation period, of the order of $\sim 10^4$~yr in \I\ at a radius
of $\sim 1600$~AU. During such episodes, the efficiency of angular
momentum transport increases by at least one order of magnitude. In
this way one might expect the inner disk to be episodically resupplied
to power the outflow. Such episodes are observed to last for a
few rotation periods, a timescale not inconsistent with the estimated
outflow kinematic age (see Table~2).  This recurrent behaviour is
also observed in simulations of magnetized disks, as a consequence of
the interaction between gravitational and magnetic instabilities
({\it Fromang et al.}, 2004a, 2004b), that can lead to a periodic variability
of the accretion rate.

Finally, it should be remembered that the energy transport in
self-gravitating disks might be dominated by wave transport, rather
than by viscous diffusion ({\it Balbus and Papaloizou}, 1999).  These
non-local effects may affect significantly the rate of mass
transfer through the disk and possibly lead to higher values of $\alpha$.
However, the evidence for these non-local phenomena remains controversial --
see {\it Lodato and Rice}~(2004, 2005), {\it Pickett et al.}~(2003), and {\it
Mej\'{\i}a et al.}~(2005) for contrasting views.

\bigskip

\centerline{\textbf{ 5. DISK LIFETIME}}
\bigskip

The radiation emitted by the central star affects the structure
and evolution of the surrounding disk in many ways. One of the most
important effects is the possibility of photo-evaporative mass
flows from the surface of the disk. These processes have been studied
extensively in the past, starting from {\it Hollenbach et al.}~(1993, 1994),
who proposed that the process of photo-evaporation of a massive disk 
could lead to the formation of an ultracompact \HII\ region.

The basic idea of the photo-evaporation is that the ionizing
ultraviolet radiation from the central star can produce a thin, hot
layer of ionized material at the surface of the disk. Far enough from
the central star, this thin layer can be hot enough to become unbound
from the star and leave the disk plane to form a photo-evaporative
outflow. The radius outside which this outflow can be produced is
obtained by equating the thermal energy of the hot gas to the
gravitational binding energy -- see Eq.~(2.1) of {\it Hollenbach et al.} (1994).
The photo-evaporation outflow rate depends sensitively on the magnitude
of the ionizing flux $\Phi$:
\begin{equation}
\dot{M}_{\rm ph}\approx 10^{-6}
\left(\frac{\Phi}{10^{47}\mbox{s}^{-1}}\right)^{1/2}
\left(\frac{M_\star}{10~M_\odot}\right)^{1/2}~\mbox{\Msun~yr$^{-1}$},
\end{equation}
where we have scaled the main parameters to values typical of a B0
star. The above estimates, based on a simple static model, 
neglect the effects of dust opacity. However, more realistic models
({\it Richling and Yorke},~1997) confirm the estimate of an
outflow rate of $\sim 10^{-6}$\Msun~yr$^{-1}$ for a typical B
star, as in the case of \I. 

Note that the photo-evaporation outflow rate estimated above is much
smaller than the typical accretion rate estimated for these systems (see
Sections.~2.3 and~4). However, the situation might be different for O stars,
where the photo-evaporation lifetime becomes comparable to the
accretion time-scale ({\it Yorke},~2004b). In order to properly assess the
interplay between viscous evolution and photo-evaporation (especially
for very high-mass stars, where the two processes occur on comparable
time-scales), we need models that incorporate both processes.  Such
models have been developed only in the context of low-mass young stars
({\it Clarke et al.}, 2001; {\it Matsuyama et al.}, 2003), but their extrapolation to
higher masses is not straightforward.

\bigskip

\centerline{\textbf{ 6. THE EFFECTS OF STELLAR COMPANIONS}}
\bigskip

A process competing with photo-evaporation in the dispersal of disk
material at distances $\sim 10^3$~AU from the central star is the tidal
stripping due to encounters with binary or cluster companions. This
process may result in the truncation of the disk at smaller radii, or
in the complete dispersal of the disk material depending on the impact
parameter of the collision. The following two limiting cases are
easy to analyse.

({\it a}\/) If the periastron of the companion is inside the primary's
disk, the interaction is highly destructive, and results in removing
all disk material external to the periastron in a single orbital
transit. A smaller circumstellar disk of radius $\sim 1/3$ of the
original periastron radius may survive or reform shortly after the
interaction, as shown by {\it Clarke and Pringle}~(1993) and {\it Hall et
al.}~(1996).  

({\it b}\/) For a wide binary system,
the maximum size of a circumstellar accretion disk is
smaller than the Roche lobe ({\it Paczynski},~1977). For stars of
comparable mass $M_1$ and $M_2$ this is a fraction
$\sim$$0.4+0.2\log(M_1/M_2)$ of the semi-major axis ({\it Paczynski},~1971). 
Under these assumptions, {\it Hollenbach et al.}~(2000) estimated a time scale
of $\sim$$2\times 10^5$~yr for dispersal of a $10^3$~AU disk in 
a cluster with stellar density $\sim 10^4$~pc$^{-3}$ and velocity dispersion
\mbox{$\sim$1~km~s$^{-1}$}.

It is clear then that a higher
frequency of wide binaries (separation $\sim 10^3$~AU) among O-type
stars might help to explain the apparent scarcity of large
circumstellar disks around these stars discussed in Section~2.  There
is in fact a possible indication of a trend for increasing degree of
multiplicity among stars of the earliest spectral types ({\it Preibisch et
al.},~2001). For example, the binary frequency in O-type stars ranges
from $\sim 40$\% ({\it Garmany et al.},~1980) to $\sim 60$\% ({\it Mason et
al.},~1998), whereas the binary frequency of B-type stars is generally lower, of
the order $\sim 14$\% ({\it McAlister et al.},~1993). Despite the large statistical
uncertainties in these estimates, the possibility of efficient tidal truncation
of circumstellar disks by binary companions among O-type stars is not
completely negligible, especially for wide binary systems with
separations comparable to the disk sizes listed in Table~2, of the order of 
$\sim 10^2$--$10^3$~AU. These systems, however, seem to be
underrepresented in the Orion Trapezium cluster relative to the main
sequence field star population ({\it McCaughrean et al.},~2000).

In addition to interactions with an orbiting exterior companion, a
circumstellar disk around a star in a dense cluster is also subject to
significant tidal forces due to the cluster's gravitational field, that
may limit the disk size and affect its evolution.  Tidal effects are
not important if the variation of the cluster's potential ${\cal V}_{\rm cl}$
over a distance of the order of the disk diameter is smaller than the
gravitational potential of the star plus disk system itself,
$2R {\rm d}{\cal V}_{\rm cl}/{\rm d}r < GM_{\rm t}/R$, implying $R^2 <
GM_{\rm t} (2{\rm d}{\cal V}_{\rm cl}/{\rm d}r)^{-1}$.

For a cluster with central density $\rho_0$ and central velocity
dispersion $v_0$, the gradient of the gravitational potential is
maximum at a distance of the order of the cluster's scale radius
$R_{\rm cl} = v_0(6/4\pi G\rho_0)^{1/2}$.  For a Plummer's
model (see {\it Spitzer},~1987), this occurs at a distance $0.71 R_{\rm cl}$
from the cluster's centre, where the enclosed mass is 19\% of the total
cluster mass $M_{\rm cl}=(4/3)\pi R_{\rm cl}^3\rho_0$.  At this radius,
$({\rm d}{\cal V}_{\rm cl}/{\rm d}r)_{\rm max}=0.38\,GM_{\rm cl}/R_{\rm cl}^2$, and the
condition on the disk radius becomes $R < 1.1\,(M_{\rm t}/M_{\rm
cl})^{1/2} R_{\rm cl}$.  Inserting appropriate numerical values, we
obtain
\begin{equation}
R < 10^4 \left({M_{\rm t}\over 10~M_\odot}\right)^{1\over 2}
\left({v_0\over 1~\mbox{km~s$^{-1}$}}\right)^{-{1\over 2}}
\left({n_\star \over 10^4~\mbox{pc$^{-3}$}}\right)^{-{1\over 4}}~\mbox{AU},
\end{equation}
where we have adopted a mean stellar mass in the cluster of 1~\Msun.
Tidal effects thus do not seem to be significant for disks of $\sim
10^3$~AU size. For clusters like the Orion Trapezium cluster, however,
with $v_0\approx 4.5$~km~s$^{-1}$ ({\it Jones and Walker},~1988) and
$n_\star\approx 4\times 10^4$~pc$^{-3}$ ({\it McCaughrean and Stauffer},
1994), tidal effects become important for $R\approx 3500$~AU,
not an unrealistic value for massive circumstellar disks, and their
contribution to the disk stability and lifetime must be properly taken
into account.

\bigskip

\centerline{\textbf{ 7. SUMMARY AND CONCLUSIONS}}
\bigskip

The way high-mass stars form is matter of debate and observations of
circumstellar disks may help to settle this issue.  The existence of
``disks'' rotating about YSOs with masses $\le20$~\Msun\ and
luminosities $\le10^4$~\Lsun\ is well established, strongly supporting a
common formation scenario across the stellar mass spectrum.  To date about
ten disks in massive stars have been found, among these IRAS\,20126+4104
being the best studied. Here one sees a Keplerian disk about a
$\sim$7~\Msun\ (proto)star associated with a precessing outflow. This
suggests the presence of a binary system, with the dominant member lying at
the center of the disk.  Using IRAS\,20126+4104 as a prototype, we have shown
that the disk is likely to be gravitationally stable and is bound to develop
spiral density
waves. These effects may be invoked to solve the problem of
transfer of material through the disk, which in a classical $\alpha$-disk is
much less than the expected accretion rate onto the star.

What is still missing is evidence of disks in more luminous objects, namely
above $\sim10^5$~\Lsun, where only huge, massive, non-equilibrium rotating
structures are detected: these we have named ``toroids''. However, absence of
evidence does not imply evidence of absence and it is still possible that O
(proto)stars are surrounded by circumstellar disks that remain undetected
because of instrumental limitations.  Indeed, the distance of these objects
is significantly larger than for B-type stars, so that observational biases
may play an important role. Also, more massive stars are expected to be
associated with richer clusters: the presence of multiple outflows/disks in
the same field may confuse the observations.  A slightly different
possibility is that disks are hidden inside the massive toroids, and one may
speculate that the latter could eventually evolve into a sample of
circumstellar disks. Alternatively, the formation of large disks might be
inhibited in O stars: we have shown that tidal interaction with companions
may be effective for this purpose, whereas disk photo-evaporation by the O
star occurs over too long a time scale.  Consequently, disks in O stars might
be truncated at small radii and hence difficult to detect with current
techniques.

In conclusion, it seems plausible that massive stars form through disk
accretion as well as low-mass ones. However, one cannot neglect the
possibility that really {\it no} disks are present in early O stars: in this
case alternative formation scenarios (coalescence, competitive accretion)
must be invoked. The advent of new generation instruments such as the Atacama
Large Millimeter Array (ALMA) with their high sensitivity and resolution is
bound to shed light on this important topic.

\textbf{ Acknowledgments.} It is a pleasure to thank Antonella Natta and
Cathie Clarke for suggestions and criticisms on the manuscript.

\bigskip

\centerline\textbf{ REFERENCES}
\bigskip
\parskip=0pt
{\small
\baselineskip=11pt

\refs Balbus S. A.  (2003) {\it Ann. Rev. Astron. Astrophys., 41}, 555-597.   
 
\refs Balbus S. A.~and Hawley J. F. (1992) {\it Astrophys. J., 400}, 610-621.   

\refs Balbus S. A.~and Papaloizou J. C.~B.  (1999) {\it Astrophys. J., 521}, 
650-658.   

\refs Barvainis R. (1984) {\it Astrophys. J., 279}, 358-362.
 
\refs Beltr{\'a}n M.~T., Cesaroni R., Neri R., Codella C., Furuya 
R.~S., Testi L., and Olmi L.  (2004) {\it Astrophys. J., 601}, L187-L190.   
 
\refs Beltr{\'a}n M.~T., Cesaroni R., Neri R., Codella C., Furuya
R.~S., Testi L., and Olmi L.  (2005) {\it Astron. Astrophys., 435}, 901-925.   
 
\refs Bernard J.~P., Dobashi K., and Momose M.  (1999) {\it Astron. Astrophys., 350}, 
197-203.   
 
\refs Bertin G.~and Lodato G.  (1999) {\it Astron. Astrophys., 350}, 694-704.   
 
\refs Beuther H., Schilke P., Sridharan T.~K., Menten K.~M., Walmsley
C.~M., and Wyrowski F.  (2002a) {\it Astron. Astrophys., 383}, 892-904.   
 
\refs Beuther H., Schilke P., Gueth F., McCaughrean M., Andersen M., 
Sridharan T.~K., and Menten K.~M.  (2002b) {\it Astron. Astrophys., 387}, 931-943.   
 
\refs Beuther H., Schilke P., and Stanke T.  (2003) {\it Astron. Astrophys., 408}, 
601-610.   
 
\refs Beuther H., Schilke P., and Gueth F.  (2004a) {\it Astrophys. J., 608}, 
330-340.   
 
\refs Beuther H., Hunter T.~R., Zhang Q., Sridharan T.~K., Zhao J.-H., 
et al.
(2004b) {\it Astrophys. J., 616}, L23-L26.   
 
\refs Beuther H., Zhang Q., Sridharan T.~K., and Chen Y.  (2005) {\it 
Astrophys. J., 628}, 800-810.   
 
\refs Beuther H., Zhang Q., Reid M. J., Hunter T.R., Gurwell M.,
et al.
(2006) {\it Astrophys. J., 636}, 323.

\refs Bik A.~and Thi W.~F.  (2004) {\it Astron. Astrophys., 427}, L13-L16.
 
\refs Birkmann S. M., Krause O., and Lemke D. (2006) {\it Astrophys. J., 637}, 380-383.
 
\refs Bonnell I. A.~and Bate M. R.  (2005) {\it Mon. Not. R. Astron. Soc., 362}, 
915-920.   
 
\refs Burrows C. J., Stapelfeldt K. R., Watson A. M., Krist J. E., Ballester G. E.,
et al.
(1996) {\it Astrophys. J., 473}, 437-451.

\refs Calvet N., Hartmann L., and Strom S.~E.  (2000) In {\it Protostars and Planets IV} (V. Mannings et al., eds.), pp. 377-399. Univ. of Arizona, Tucson.
 
\refs Cesaroni R. (2005a) {\it Astrophys. Space Sci., 295}, 5-17.   
 
\refs Cesaroni R. (2005b) In {\it Massive Star Birth: A Crossroads of Astrophysics}, {\it IAU Symposium 227} (R. Cesaroni et al., eds.), pp. 59-69. Cambridge Univ., Cambridge.

\refs Cesaroni R., Olmi L., Walmsley C.~M., Churchwell E., and Hofner
P.  (1994) {\it Astrophys. J., 435}, L137-L140.   
 
\refs Cesaroni R., Felli M., Testi L., Walmsley C.~M., and Olmi L.  
(1997) {\it Astron. Astrophys., 325}, 725-744.   
 
\refs Cesaroni R., Hofner P., Walmsley C.~M., and Churchwell E.  (1998) 
{\it Astron. Astrophys., 331}, 709-725.   
 
\refs Cesaroni R., Felli M., Jenness T., Neri R., Olmi L., Robberto M.,
Testi L., and Walmsley C.~M.  (1999a) {\it Astron. Astrophys., 345}, 949-964.
 
\refs Cesaroni R., Felli M., and Walmsley C.~M.  (1999b) {\it Astron. Astrophys. Suppl., 
136}, 333-361.
 
\refs Cesaroni R., Neri R., Olmi L., Testi L., Walmsley C.~M., and 
Hofner P.  (2005) {\it Astron. Astrophys., 434}, 1039-1054.   
 
\refs Chini R., Hoffmeister V., Kimeswenger S., Nielbock
M., N{\"u}rnberger D., Schmidtobreick L., and Sterzik M. 
(2004) {\it Nature, 429}, 155-157.   
 
\refs Churchwell E., Walmsley C.~M., and Cesaroni R.  (1990) {\it Astron. Astrophys. Suppl., 
83}, 119-144.   
 
\refs Clarke C.~J.~and Pringle J.~E.  (1993) {\it Mon. Not. R. Astron. Soc., 261}, 190-202.   

\refs Clarke C.~J., Gendrin A., and Sotomayor M.  (2001) {\it Mon. Not. R. Astron. Soc., 
328}, 485-491.   
 
\refs Codella C., Lorenzani A., Gallego A.~T., Cesaroni R., and 
Moscadelli L.  (2004) {\it Astron. Astrophys., 417}, 615-624.   
 
\refs Cohen R.~J., Baart E.~E., and Jonas J.~L.  (1988) {\it Mon. Not. R. Astron. Soc., 
231}, 205-227.   
 
\refs De Buizer J.~M.  (2003) {\it Mon. Not. R. Astron. Soc., 341}, 277-298.   
 
\refs De Buizer J. M.~and Minier V. (2005) {\it Astrophys. J., 628}, 
L151-L154.   
 
\refs Dent W.~R.~F., Little L.~T., Kaifu N., Ohishi M., and Suzuki S.  
(1985) {\it Astron. Astrophys., 146}, 375-380.   
 
\refs Edris K.~A., Fuller G.~A., Cohen R.~J., and Etoka S.  (2005) {\it 
Astron. Astrophys., 434}, 213-220.   
 
\refs Elitzur M. (1992) {\it Ann. Rev. Astron. Astrophys., 30}, 75-112.   
 
\refs Estalella R., Mauersberger R., Torrelles J. M., Anglada
G., G\'omez J. F., Lopez R., and Muders D. (1993) {\it Astrophys. J., 
419}, 698-706.
 
\refs Fontani F., Cesaroni R., Testi L., Molinari S., Zhang Q., Brand
J., and Walmsley C.~M.  (2004) {\it Astron. Astrophys., 424}, 179-195.  

\refs Fromang S., Balbus S. A., and De Villiers
J.-P. (2004a) {\it Astrophys. J., 616}, 357-363.   
 
\refs Fromang S., Balbus S. A., Terquem C., and De 
Villiers J.-P. (2004b) {\it Astrophys. J., 616}, 364-375.   
 
\refs Fuente A., Rodr{\'{\i}}guez-Franco A., Testi L., Natta A., 
Bachiller R., and Neri R.  (2003) {\it Astrophys. J., 598}, L39-L42.   
 
\refs Fuller G.~A., Zijlstra A.~A., and Williams S.~J.  (2001) {\it Astrophys. J., 
555}, L125-L128.   
 
\refs Galli D. and Shu F. H.  (1993a) {\it Astrophys. J., 417}, 243-258.

\refs Galli D. and Shu F. H.  (1993b) {\it Astrophys. J., 417}, 220-242.

\refs Gammie C. F.  (2001) {\it Astrophys. J., 553}, 174-183.   
 
\refs Garmany C.~D., Conti P.~S., and Massey P.  (1980) {\it Astrophys. J., 242}, 
1063-1076.   

\refs Genzel R. and Stutzki J. (1989) {\it Ann. Rev. Astron. Astrophys., 27}, 
41-85.   

\refs Gibb A.~G., Hoare M.~G., Mundy L.~G., and Wyrowski F.  (2004a) In {\it Star Formation at High Angular Resolution},
{\it IAU Symposium 221}, 425-430. Kluwer/Springer, Dordrecht.
 
\refs Gibb A.~G., Wyrowski F., and Mundy L.~G.  (2004b) {\it Astrophys. J., 616}, 
301-318.   
 
\refs G{\'o}mez J. F., Sargent A. I., Torrelles J. M., 
Ho P. T.~P., Rodr{\'{\i}}guez L. F., Cant{\'o} J., and Garay
G. (1999) {\it Astrophys. J., 514}, 287-295.   
 
\refs Goodman A.~A., Benson P.~J., Fuller G.~A., and Myers P.~C.  
(1993) {\it Astrophys. J., 406}, 528-547.   
 
\refs Greenhill L.~J., Reid M.~J., Chandler C.~J., Diamond P.~J., and 
Elitzur M.  (2004) In {\it Star Formation at High Angular Resolution}, {\it IAU Symposium 221} (M. Burton et al., eds.), pp. 155-160. Kluwer/Springer, Dordrecht.
 
\refs Guilloteau S., Dutrey A., and Simon M.  (1999) {\it Astron. Astrophys., 348}, 
570-578.   
 
\refs Hall S.~M., Clarke C.~J., and Pringle J.~E.  (1996) {\it Mon. Not. R. Astron. Soc., 
278}, 303-320.   

\refs Hasegawa T. I.~and Mitchell G. F.  (1995) {\it Astrophys. J., 
451}, 225-237.
 
\refs Hillenbrand L. A., Carpenter J.~M., and Skrutskie M.~F.
(2001) {\it Astrophys. J., 547}, L53-L56.
 
\refs Hofner P., Cesaroni R., Rodr{\'{\i}}guez L.~F., and Mart{\'{\i}}
J.  (1999) {\it Astron. Astrophys., 345}, L43-L46.   
 
\refs Hollenbach D., Johnstone D., and Shu F.  (1993) {\it ASPC, 35}, 
26-34.
 
\refs Hollenbach D., Johnstone D., Lizano S., and Shu F. 
(1994) {\it Astrophys. J., 428}, 654-669.   
 
\refs Hollenbach D.~J., Yorke H.~W., and Johnstone D.  (2000) In {\it Protostars and Planets IV} (V.  Mannings et al., eds.), pp. 401-428. Univ. of Arizona, Tucson.
 
\refs Hunter T.~R., Testi L., Zhang Q., and Sridharan T.~K.  (1999) 
{\it Astron. J., 118}, 477-487.   
 
\refs Hutawarakorn B.~and Cohen R.~J.  (1999) {\it Mon. Not. R. Astron. Soc., 303}, 845-854.  

\refs Jiang Z., Tamura M., Fukagawa M., Hough J., Lucas P.,
Suto H., Ishii M., and Yang J.
(2005) {\it Nature, 437}, 112-115. 

\refs Jijina J. and Adams F. C.  (1996) {\it Astrophys. J., 462}, 874-887.
 
\refs Jones B.~F.~and Walker M. F.  (1988) {\it Astron. J., 95}, 1755-1782.   

\refs Johnson B. M.~and Gammie C. F.  (2003) {\it Astrophys. J., 597}, 
131-141.   
 
\refs Kahn F.~D.  (1974) {\it Astron. Astrophys., 37}, 149-162.   
 
\refs Keto E. R., Ho P. T.~P., and Haschick A. D.  (1988) {\it 
Astrophys. J., 324}, 920-930.   
 
\refs K\"onigl A.~and Pudritz R.~E.  (2000) In {\it Protostars and Planets IV} (V.  Mannings et al., eds.), pp. 759-787. Univ. of Arizona, Tucson.

\refs Kramer C., Alves J., Lada C., Lada E., Sievers A., Ungerechts
H., and Walmsley M.  (1998) {\it Astron. Astrophys., 329}, L33-L36.   
 
\refs Krolik J. H.  (1999) {\it Active galactic nuclei: from the central black hole to the galactic environment}, Princeton University.
 
\refs Krumholz M. R., McKee C. F., and Klein R. I.  
(2005) {\it Astrophys. J., 618}, L33-L36.   
 
\refs Laughlin G. and Bodenheimer P. (1994) {\it Astrophys. J., 436}, 
335-354.   
 
\refs Laughlin G. and Korchagin V. (1996) 
{\it Astrophys. J., 460}, 855-868.
 
\refs Liu S.-Y. and the SMA Team (2005) In {\it Massive Star Birth: A Crossroads of Astrophysics}, {\it IAU Symposium 227} (R. Cesaroni et al., eds.), pp. 47-52. Cambridge Univ., Cambridge.

\refs Lodato G.~and Rice W.~K.~M.  (2004) {\it Mon. Not. R. Astron. Soc., 351}, 630-642.   
 
\refs Lodato G.~and Rice W.~K.~M.  (2005) {\it Mon. Not. R. Astron. Soc., 358}, 1489-1500.   
 
\refs MacLeod G.~C.~and Gaylard M.~J.  (1992) {\it Mon. Not. R. Astron. Soc., 256}, 519-527.  

\refs Mart\'{\i} J., Rodr\'{\i}guez L.~F., and Reipurth B.  (1993) {\it Astrophys. J., 416}, 208-217.
 
\refs Mason B. D., Henry T. J., Hartkopf W. I., Ten Brummelaar T., and
Soderblom D. R.  (1998) {\it Astron. J., 116}, 2975-2983. 

\refs Mathieu R. D., Ghez A. M., Jensen E. L. N., and Simon M. (2000) In {\it Protostars and Planets IV} (V.  Mannings et al., eds.), pp. 703-730. Univ. of Arizona, Tucson.
 
\refs Matsuyama I., Johnstone D., and Hartmann L. (2003) {\it Astrophys. J., 
582}, 893-904.   
 
\refs McAlister H. A., Mason B. D., Hartkopf W. I., and 
Shara M. M.  (1993) {\it Astron. J., 106}, 1639-1655.   

\refs McCaughrean M. J.~and Stauffer J. R.  (1994) {\it Astron. J., 108}, 
1382-1397.   

\refs McCaughrean M.~J., Stapelfeldt K.~R., and Close L.~M.  (2000) In {\it 
Protostars and Planets IV} (V.  Mannings et al., eds.), pp. 485-507. Univ. of Arizona, Tucson.

\refs McKee C. F.~and Tan J. C.  (2003) {\it Astrophys. J., 585}, 
850-871.   
 
\refs Mej{\'{\i}}a A. C., Durisen R. H., Pickett M. K., and 
Cai K. (2005) {\it Astrophys. J., 619}, 1098-1113.   
 
\refs Mestel L.  (1963) {\it Mon. Not. R. Astron. Soc., 126}, 553-575.

\refs Minier V., Booth R.~S., and Conway J.~E.  (1998) {\it Astron. Astrophys., 336}, 
L5-L8.   
 
\refs Minier V., Booth R.~S., and Conway J.~E.  (2000) {\it Astron. Astrophys., 362}, 
1093-1108.   
 
\refs Mitchell G. F., Hasegawa T. I., and Schella J. 
(1992) {\it Astrophys. J., 386}, 604-617.   
 
\refs Molinari S., Brand J., Cesaroni R., and Palla F.  (1996) {\it 
Astron. Astrophys., 308}, 573-587.   

\refs Molinari S., Brand J., Cesaroni R., Palla F., and Palumbo
G.~G.~C.  (1998) {\it Astron. Astrophys., 336}, 339-351.   

\refs Moscadelli L., Cesaroni R., and Rioja M.~J.  (2000) {\it Astron. Astrophys., 
360}, 663-670.   
 
\refs Moscadelli L., Cesaroni R., and Rioja M.~J.  (2005) {\it Astron. Astrophys., 
438}, 889-898.   
 
\refs Mundy L.~G., Looney L.~W., and Welch W.~J.  (2000) In {\it Protostars and Planets IV} (V.  Mannings et al., eds.), pp. 355-376. Univ. of Arizona, Tucson.
 
\refs Nakano T., Hasegawa T., Morino J.-I., and Yamashita, 
T. (2000) {\it Astrophys. J., 534}, 976-983.   

\refs Natta A., Grinin V., and Mannings V. (2000) In {\it Protostars and Planets IV} (V.  Mannings et al., eds.), pp. 559-587. Univ. of Arizona, Tucson.
 
\refs Norris R.~P., Byleveld S.~E., Diamond P.~J., Ellingsen S.~P., Ferris R.~H.,
et al.
(1998) {\it Astrophys. J., 508}, 275-285.   
 
\refs O'Dell C.~R.  (2001) {\it Astron. J., 122}, 2662-2667.   
 
\refs O'Dell C.~R.~and Wen Z. (1994) {\it Astrophys. J., 436}, 194-202.   

\refs Olmi L., Cesaroni R., Hofner P., Kurtz S., Churchwell E., and 
Walmsley C.~M.  (2003) {\it Astron. Astrophys., 407}, 225-235.   
 
\refs Osterloh M., Henning Th., and Launhardt R.  (1997) {\it Astrophys. J. Suppl., 
110}, 71-114.
 
\refs Paczynski B.  (1971) {\it Ann. Rev. Astron. Astrophys., 9}, 183-208.

\refs Paczynski B.  (1977) {\it Astrophys. J., 216}, 822-826.   

\refs Palla F. and Stahler S. W.  (1993) {\it Astrophys. J., 418}, 414-425.
 
\refs Palla F., Brand J., Comoretto G., Felli M., and Cesaroni R.  
(1991) {\it Astron. Astrophys., 246}, 249-263.   
 
\refs Patel N. A., Curiel S., Sridharan T. K., Zhang Q., Hunter T. R.,
et al.
(2005) {\it Nature, 437}, 109-111. 

\refs Pestalozzi M. R., Elitzur M., Conway J. E., and Booth
R. S.  (2004) {\it Astrophys. J., 603}, L113-L116.   
 
\refs Phillips C.~J., Norris R.~P., Ellingsen S.~P., and McCulloch
P.~M.  (1998) {\it Mon. Not. R. Astron. Soc., 300}, 1131-1157.   
 
\refs Pickett B. K., Cassen P., Durisen R. H., and Link
R. (2000) {\it Astrophys. J., 529}, 1034-1053.   
 
\refs Pickett B. K., Mej{\'{\i}}a A. C., Durisen R. H., 
Cassen P. M., Berry D. K., and Link R. P.  (2003) {\it Astrophys. J., 
590}, 1060-1080.   
 
\refs Plume R., Jaffe D.~T., and Evans N. J., II (1992) {\it Astrophys. J. Suppl., 
78}, 505-515.   
 
\refs Preibisch Th., Weigelt G., and Zinnecker H.  (2001) In {\it The Formation of Binary Stars}, {\it IAU Symposium 200} (H. Zinnecker and R. Mathieu, eds.), pp.~69-78. Kluwer/Springer, Dordrecht.

\refs Preibisch T., Balega Y.~Y., Schertl D., and Weigelt G.  (2003) 
{\it Astron. Astrophys., 412}, 735-743.   
 
\refs Rice W.~K.~M., Lodato G., and Armitage P.~J.  (2005) {\it Mon. Not. R. Astron. Soc., 346}, L56-L60.
 
\refs Richer J.~S., Shepherd D.~S., Cabrit S., Bachiller R., and 
Churchwell E.  (2000) In {\it Protostars and Planets IV} (V. Mannings et al., eds.), pp. 867-894. Univ. of Arizona, Tucson.

\refs Richling S.~and Yorke H.~W.  (1997) {\it Astron. Astrophys., 327}, 317-324.   
 
\refs Rodr\'{\i}guez L. F. (2005) In {\it Massive Star Birth: A Crossroads of Astrophysics}, {\it IAU Symposium 227} (R. Cesaroni et al., eds.), pp. 120-127, Cambridge Univ., Cambridge.

\refs Sako S., Yamashita T., Kataza H., Miyata T., Okamoto Y. K.,
et al.
(2005) {\it Nature, 434}, 995-998.   
 
\refs Sandell G., Wright M., and Forster J. R.  (2003) 
{\it Astrophys. J., 590}, L45-L48.   
 
\refs Schreyer K., Henning Th., van der Tak F.~F.~S., Boonman A.~M.~S., 
and van Dishoeck E.~F.  (2002) {\it Astron. Astrophys., 394}, 561-583.   
 
\refs Schreyer K., Semenov D., Henning Th., and Forbrich J. (2006) {\it Astrophys. J., 637}, L129-L132

\refs Scoville N., Kleinmann S.~G., Hall D.~N.~B., and Ridgway S.~T.  
(1983) {\it Astrophys. J., 275}, 201-224.   

\refs Shakura N.~I.~and Sunyaev R.~A.  (1973) {\it Astron. Astrophys., 24}, 337-355.   
 
\refs Shepherd D.~S. (2003) In {\it Galactic Star Formation Across the
Stellar Mass Spectrum} (J. M. De Buizer and N. S. van der Bliek, eds.),
pp. 333-344. Astronomical Society of the Pacific, San Francisco.

\refs Shepherd D.~S.~and Churchwell E.  (1996a) {\it Astrophys. J., 457}, 267-276.
 
\refs Shepherd D.~S.~and Churchwell E.  (1996b) {\it Astrophys. J., 472}, 225-239.
 
\refs Shepherd D.~S.~and Kurtz S.~E.  (1999) {\it Astrophys. J., 523}, 690-700.   
 
\refs Shepherd D.~S., Yu K.~C., Bally J., and Testi L.  (2000) {\it 
Astrophys. J., 535}, 833-846.   
 
\refs Shepherd D.~S., Claussen M.~J., and Kurtz S.~E.  (2001) {\it Science,
292}, 1513-1518.   
 
\refs Shu F.~H., Najita J.~R., Shang H., and Li Z.-Y.  (2000) In {\it 
Protostars and Planets IV} (V. Mannings et al., eds.), pp. 789-813. Univ. of Arizona, Tucson.

\refs Simon M., Dutrey A., and Guilloteau S.  (2000) {\it Astrophys. J., 545}, 
1034-1043.   
 
\refs Sollins P. K., Zhang Q., Keto E., and Ho P. T.~P.  
(2005a) {\it Astrophys. J., 624}, L49-L52.   
 
\refs Sollins P. K., Zhang Q., Keto E., and Ho P. T.~P.  
(2005b) {\it Astrophys. J., 631}, 399-410.   
 
\refs Spitzer L. (1987) {\it Dynamical Evolution of Globular Clusters},
Princeton University, Princeton, New Jersey.

\refs Sridharan T.~K., Beuther H., Schilke P., Menten K.~M., and 
Wyrowski F.  (2002) {\it Astrophys. J., 566}, 931-944.   
 
\refs Sridharan T.~K., Williams S.~J., and Fuller G.~A.  (2005) {\it 
Astrophys. J., 631}, L73-L76.   

\refs Stahler S.~W., Palla F., and Ho P.~T.~P.  (2000) In {\it Protostars and Planets IV} (V. Mannings et al., eds.), pp. 327-351. Univ. of Arizona, Tucson.

\refs Su Y.-N., Zhang Q., and Lim J. (2004) {\it Astrophys. J., 604}, 
258-271.   

\refs Tan J. C.~and McKee C. F.  (2004) {\it Astrophys. J., 603}, 
383-400.   
 
\refs Terebey S., Shu F.~H., and Cassen P.  (1984) {\it Astrophys. J., 286}, 
529-551.   
 
\refs Tofani G., Felli M., Taylor G.~B., and Hunter T.~R.  (1995) {\it 
Astron. Astrophys. Suppl., 112}, 299-346.
 
\refs Toomre A.  (1964) {\it Astrophys. J., 139}, 1217-1238.   
 
\refs Torrelles J. M., G{\'o}mez J. F., Garay G., 
Rodr{\'{\i}}guez L. F., Curiel S., Cohen R.~J., and Ho P. 
T.~P.  (1998) {\it Astrophys. J., 509}, 262-269.   
 
\refs van der Tak F.~F.~S.~and Menten K.~M.  (2005) {\it Astron. Astrophys., 437}, 
947-956.   
 
\refs van der Tak F. F. S., Walmsley C. M., Herpin F., and Ceccarelli C. (2006) {\it Astron. Astrophys., 447}, 1011-1025.

\refs Walker C. K., Adams F. C., and Lada C. J.  (1990) 
{\it Astrophys. J., 349}, 515-528.   
 
\refs Wilking B.~A., Blackwell J.~H., Mundy L.~G., and Howe J.~E.  
(1989) {\it Astrophys. J., 345}, 257-264.   
 
\refs Wilking B. A., Blackwell J. H., and Mundy L. G.  (1990) 
{\it Astron. J., 100}, 758-770.   
 
\refs Wilner D.~J.~and Lay O.~P.  (2000) In {\it Protostars and Planets IV} (V. Mannings et al., eds.), pp. 509-532. Univ. of Arizona, Tucson.
 
\refs Wolfire M. G.~and Cassinelli J. P.  (1987) {\it Astrophys. J., 319}, 
850-867.   
 
\refs Wright M.~C.~H., Plambeck R.~L., Mundy L.~G., and Looney L.~W.  
(1995) {\it Astrophys. J., 455}, L185-L188.
 
\refs Yao Y., Ishii M., Nagata T., Nakaya H., and 
Sato S. (2000) {\it Astrophys. J., 542}, 392-399.  
 
\refs Yorke H.~W. (2004a) In {\it Star Formation at High Angular Resolution}, {\it IAU Symposium 221} (M. Burton et al., eds.), pp.~141-152. Kluwer/Springer, Dordrecht.
 
\refs Yorke H.~W. (2004b) {\it Rev. Mex. Astron. Astrofis. Ser. Conf., 22}, 42-45.  
 
\refs Yorke H. W.~and Sonnhalter C. (2002) {\it Astrophys. J., 569}, 846-862.  
 
\refs Zhang Q. (2005) In {\it Massive Star Birth: A Crossroads of Astrophysics}, {\it IAU Symposium 227} (R. Cesaroni et al., eds.), pp. 135-144. Cambridge Univ., Cambridge.

\refs Zhang Q., Ho P. T.~P., and Ohashi N. (1998a) {\it Astrophys. J., 
494}, 636-656.  
 
\refs Zhang Q., Hunter T. R., and Sridharan T.~K. (1998b) {\it 
Astrophys. J., 505}, L151-L154.  
 
\refs Zhang Q., Hunter T. R., Sridharan T.~K., and Cesaroni
R. (1999) {\it Astrophys. J., 527}, L117-L120.  
 
\refs Zhang Q., Hunter T.~R., Brand J., Sridharan T.~K., Molinari S.,
et al.
(2001) {\it Astrophys. J., 552}, L167-L170.  
 
\refs Zhang Q., Hunter T. R., Sridharan T.~K., and Ho Paul T.~P.
(2002) {\it Astrophys. J., 566}, 982-992.  
 
\refs Zhang Q., Hunter T.~R., Brand J., Sridharan T.~K., Cesaroni R.,
et al.
(2005) {\it Astrophys. J., 625}, 864-882.  


\end{document}